\documentclass{article}

\usepackage[french]{babel}
\usepackage{geometry}    
\usepackage[T1]{fontenc} 
\usepackage{amsfonts}    
\usepackage{amssymb}     
\usepackage[all]{xy}     
\usepackage{amsmath}     

\newcommand{\go}[1]{\mathfrak {#1}}
\newcommand{\rrbr}{]\!]}
\newcommand{\llbr}{[\![}

\newtheorem{The}{Theorem}
\newtheorem{Def}[The]{Definition}
\newtheorem{Lem}[The]{Lemma}
\newtheorem{Pro}[The]{Proposition}
\newtheorem{Cor}[The]{Corollary}
\newtheorem{Rq}[The]{Remark}
\newtheorem{Ex}[The]{Example}
\newtheorem{Conj}[The]{Conjecture}

\begin{document}

\small

\begin{center}
\Large{Poisson Homology in Degree $0$ for some Rings of Symplectic Invariants}\vspace{.25cm}\\
\end{center}

\begin{center}
Fr\'ed\'eric BUTIN\footnote{Universit\'e de Lyon, Universit\'e
Lyon 1, CNRS, UMR5208, Institut
Camille Jordan,\\
43 blvd du 11 novembre 1918, F-69622 Villeurbanne-Cedex, France\\
email: butin@math.univ-lyon1.fr
}\vspace{.5cm}\\
\end{center}

\begin{small}
\noindent\textbf{\textsc{Abstract}}\\
Let $\go{g}$ be a finite-dimensional semi-simple Lie algebra,
$\go{h}$ a Cartan subalgebra of $\go{g}$, and $W$ its Weyl group.
The group $W$ acts diagonally on $V:=\go{h}\oplus\go{h}^*$, as
well as on $\mathbb{C}[V]$. The purpose of this article is to
study the Poisson homology of the algebra of invariants
$\mathbb{C}[V]^W$ endowed with the standard symplectic bracket.\\
To begin with, we give general results about the Poisson homology
space in degree $0$, denoted by $HP_0(\mathbb{C}[V]^W)$, in the
case where $\go{g}$ is of type $B_n-C_n$ or $D_n$, results which
support Alev's conjecture. Then we are focusing the interest on
the particular cases of ranks $2$ and $3$, by computing the
Poisson homology space in degree $0$ in the cases where $\go{g}$
is of type $B_2$ ($\go{so}_5$), $D_2$ ($\go{so}_4$), then $B_3$
($\go{so}_7$), and $D_3=A_3$ ($\go{so}_6\simeq\go{sl}_4$). In
order to do this, we make use of a functional equation introduced
by Y. Berest, P. Etingof and V. Ginzburg. We recover, by a
different method, the result established by J. Alev and L. Foissy,
according to which the dimension of $HP_0(\mathbb{C}[V]^W)$ equals
$2$ for $B_2$. Then we calculate the dimension of this space and
we show that it is equal to 1 for $D_2$. We also calculate it
for the rank $3$ cases, we show that it is equal to $3$ for $B_3-C_3$ and $1$ for $D_3=A_3$.\\
\end{small}

\begin{small}
\noindent\textbf{\textsc{Key-words}}\\
Alev's conjecture; Pfaff; Poisson homology; Weyl group; invariants; Berest-Etingof-Ginzburg equation.\\
\end{small}

\section{\textsf{Introduction}}

\noindent Let $G$ be a finite subgroup of the symplectic group
$\mathbf{Sp}(V)$, where $V$ is a $\mathbb{C}-$vector space of
dimension~$2n$. Then the algebra of polynomial functions on $V$,
denoted by $\mathbb{C}[V]$, is a Poisson algebra for the standard
symplectic bracket, and as $G$ is a subgroup of the symplectic
group, the algebra of invariants, denoted by
$\mathbb{C}[V]^G$, is also a Poisson algebra.\\

\noindent Several articles were devoted to the computation of
Poisson homology and cohomology of the algebra of
invariants~$\mathbb{C}[V]^G$. In particular, Y. Berest, P. Etingof
and V. Ginzburg, in \cite{BEG04}, prove that the $0-$th space of
Poisson homology of $\mathbb{C}[V]^G$ is finite-dimensional.\\
After their works \cite{AL98} and \cite{AL98}, J. Alev, M. A.
Farinati, T. Lambre and A. L. Solotar establish a fundamental
result in \cite{AFLS00}: they compute all the spaces of Hochschild
homology and cohomology of $A_n(\mathbb{C})^G$ for every finite
subgroup $G$ of $\mathbf{Sp}_{2n}\mathbb{C}$.\\
Besides, J. Alev and L. Foissy show in \cite{AF06} that the
dimension of the Poisson homology space in degree $0$ of
$\mathbb{C}[\go{h}\oplus\go{h}^*]^W$ is equal to the one of the
Hochschild homology space in degree $0$ of $A_2(\mathbb{C})^W$,
where $\go{h}$ is a Cartan subalgebra of a semi-simple Lie algebra
of rank $2$ with Weyl group $W$.\\

\noindent In the following, given a finite-dimensional semi-simple
Lie algebra $\go{g}$, a Cartan subalgebra $\go{h}$ of $\go{g}$,
and its Weyl group $W$, we are interested in the Poisson homology
of $\mathbb{C}[V]^G$ in the case where $V$ is the symplectic space
$V:=\go{h}\oplus\go{h}^*$ and $G:=W$.\\
The group $W$ acts diagonally on $V$, this induces an action of
$W$ on $\mathbb{C}[V]$. We denote by $\mathbb{C}[V]^W$ the algebra
of invariants under this action. Endowed with the standard
symplectic bracket, this algebra is a Poisson algebra. The action
of the group $W$ on $V$ also induces an action of $W$ on
the Weyl algebra $A_n(\mathbb{C})$.\\
To begin with, we will give general results about the Poisson
homology space in degree $0$ of $\mathbb{C}[V]^W$ for the types
$B_n-C_n$ and $D_n$, results which support Alev's conjecture and
establish a framework for a possible proof. Sections
2.3, 2.4 and 2.5 contain the main results of this study.\\
Next we will use these results in order to completely calculate
the Poisson homology space in degree $0$, denoted by
$HP_0(\mathbb{C}[V]^W)$, in the case where $\go{g}$ is $\go{so}_5$
(i.e. $B_2$) --- so we recover, by a different method, the result
established by J. Alev and L. Foissy for $\go{so}_5$ in the
article \cite{AF06}, namely $\dim\,HP_0(\mathbb{C}[V]^W)=2$ ---
then in the case where $\go{g}$ is $\go{so}_4$ (i.e.
$D_2=A_1\times A_1$) by showing that
$\dim\,HP_0(\mathbb{C}[V]^W)=1$. Finally we will prove
the important property for rank $3$:\\

\begin{Pro}(Poisson homology in degree $0-$ for $\go{g}$ of rank $3$)$\\$
Let $HP_0(\mathbb{C}[V]^W)$ be the Poisson homology space in
degree $0$ of $\mathbb{C}[V]^W$ and
$HH_0\left(A_n(\mathbb{C})^W\right)$ the Hochschild homology space in degree $0$ of $A_n(\mathbb{C})^W$.\\
For $\go{g}$ of type $B_3$ ($\go{so}_7$), we have $\dim\,HP_0\left(\mathbb{C}[V]^W\right)=\dim\,HH_0\left(A_n(\mathbb{C})^W\right)=3$.\\
For $\go{g}$ of type $D_3=A_3$ ($\go{so}_6\simeq\go{sl}_4$), we have $\dim\,HP_0\left(\mathbb{C}[V]^W\right)=\dim\,HH_0\left(A_n(\mathbb{C})^W\right)=1$.\\
\end{Pro}

\noindent In order to carry out the computation we will use the
functional equation given in the article \cite{BEG04} quoted above.\\

\section{\textsf{Results about $B_n-C_n$ and $D_n$}}

\noindent As indicated above, we are interested in the Poisson
homology of $\mathbb{C}[\go{h}\oplus\go{h}^*]^W$, where $\go{h}$
is a Cartan subalgebra of a finite-dimensional semi-simple Lie
algebra $\go{g}$, and $W$ its Weyl group. We will study the types
$B_n$ and $D_n$. Recall that the root system of type $C_n$ is dual
to the root system of type $B_n$. So their Weyl groups are
isomorphic,
and the study of the case $C_n$ is reduced to the study of the case $B_n$.\\

\subsection{\textsf{Definitions and notations}}

\noindent $\bullet$ Set $S:=\mathbb{C}[\mathbf{x},\
\mathbf{y}]=\mathbb{C}[x_1,\dots,\ x_n,\ y_1,\dots,\ y_n]$.\\
For $m\in \mathbb{N}$, we denote by $S(m)$
the elements of $S$ of degree $m$.\\
For $B_n$, we have $W=(\pm 1)^n\rtimes \go{S}_n=(\pm 1)^n\cdot \go{S}_n$ (permutations of the variables and sign changes of the variables).\\
For $D_n$, we have $W=(\pm 1)^{n-1}\rtimes \go{S}_n=(\pm 1)^{n-1}\cdot \go{S}_n$ (permutations of the variables and sign changes of an even number of variables).\\
Every element $(a_1,\ \dots,\ a_n)\in(\pm 1)^n$ is identified with
the diagonal matrix $Diag(a_1,\ \dots,\ a_n)$, and every element
$\sigma\in\go{S}_n$ is identified with the matrix
$(\delta_{i,\sigma(j)})_{(i,j)\in\llbr 1,\ n\rrbr}$. We will
denote by $s_j$ the $j-$th sign change, i.e.
\begin{center}
$s_j(x_k)=x_k$ if $k\neq j$, $s_j(x_j)=-x_j$, and $s_j(y_k)=y_k$ if $k\neq j$, $s_j(y_j)=-y_j$.\\
\end{center} As for the elements of $(\pm 1)^{n-1}$, they are
identified with the matrices of the form
\begin{center}
$Diag((-1)^{i_1},\ (-1)^{i_1+i_2},\ (-1)^{i_2+i_3},\ \dots,\
(-1)^{i_{n-2}+i_{n-1}},\ (-1)^{i_{n-1}}),$
\end{center}
with $i_k\in\{0,\ 1\}$.  We will denote by $s_{i,j}$ the sign
change of the variables of indices $i$ and $j$.\\
So, all these elements are in $\mathbf{O}_n\mathbb{C}$, and by
identifying $g\in W$ with $\left(
\begin{array}{cc}
  g & 0 \\
  0 & g \\
\end{array}
\right)$, we obtain $W\subset \mathbf{Sp}_{2n}\mathbb{C}$.\\

\noindent $\bullet$ The (right) action of $W$ on $S$ is defined
for $P\in S$ and $g\in W$ by
$$g\cdot P(\mathbf{x},\ \mathbf{y}):=P\left(\sum_{j=1}^n g_{1j}x_j,\ \dots,\ \sum_{j=1}^n g_{nj}x_j,\ \sum_{j=1}^n g_{1j}y_j,\ \dots,\ \sum_{j=1}^n g_{nj}y_j\right).$$
So we have $h\cdot(g\cdot P)=(gh)\cdot P$.\\
In the particular case where $\sigma\in\go{S}_n$, we have
$\sigma\cdot P(\mathbf{x},\ \mathbf{y})=P(x_{\sigma^{-1}(1)},\
\dots,\ x_{\sigma^{-1}(n)},\
y_{\sigma^{-1}(1)},\ \dots,\ y_{\sigma^{-1}(n)})$.\\

\noindent $\bullet$ On $S$, we define the Poisson bracket
$$\{P,\ Q\}:=\langle \nabla P,\ \nabla
Q\rangle=\nabla P\cdot(J\,\nabla
Q)=\nabla_\mathbf{x}P\cdot\nabla_\mathbf{y}Q-\nabla_\mathbf{y}P\cdot\nabla_\mathbf{x}Q,$$
where $\langle\cdot,\,\cdot\rangle$ is the standard symplectic
product, associated to the matrix $J=\left(\begin{array}{cc}
  0 & I_n \\
  -I_n & 0 \\
\end{array}
\right)$.\\
As the group $W$ is a subgroup of
$\mathbf{Sp}(\langle\cdot,\,\cdot\rangle)=\mathbf{Sp}_{2n}\mathbb{C}$,
the algebra of invariants $S^W$ is a Poisson algebra for the bracket defined above.\\

\noindent $\bullet$ We define the Reynolds operator as the linear
map $R_n$ from $S$ to $S$ determined by
$$R_n(P)=\frac{1}{|W|}\sum_{g\in W}g\cdot P.$$
We set $A=\mathbb{C}[\mathbf{z},\
\mathbf{t}]=\mathbb{C}[z_1,\dots,\ z_n,\ t_1,\dots,\ t_n]$ and
$S':=A[\mathbf{x},\ \mathbf{y}]$, and we extend the map $R_n$
as a $A-$linear map from $S'$ to $S'$.\\

\begin{Rq}$\\$
In the case of $B_n$, every element of $S^W$ has an even degree.
(It is false for $D_n$).
\end{Rq}

\subsection{\textsf{Poisson homology}}

\noindent $\bullet$ Let $A$ be a Poisson algebra. We  denote by
$\Omega^p(A)$ the $A-$module of Kähler differentials, i.e. the
vector space spanned by the elements of the form
$F_0\,dF_1\wedge\dots\wedge dF_p$, where the $F_j$ belong to $A$,
and $d:\Omega^p(A)\rightarrow \Omega^{p+1}(A)$ is the De
Rham differential.\\
We consider the complex $$\xymatrix{\dots \ar@{->}[r]^{\partial_5}
& \Omega^4(A) \ar@{->}[r]^{\partial_4}
 & \Omega^3(A) \ar@{->}[r]^{\partial_3}
 & \Omega^2(A) \ar@{->}[r]^{\partial_2}
 & \Omega^1(A) \ar@{->}[r]^{\partial_1}
 & \Omega^0(A) }$$
with Brylinsky-Koszul boundary operator $\partial_p$ (See
\cite{B88}):
$$\begin{array}{rcl}
   \partial_p(F_0\,dF_1\wedge\dots\wedge
dF_p) & = & \displaystyle{\sum_{j=1}^p
(-1)^{j+1}\{F_0,\,F_j\}\,dF_1\wedge\dots\wedge
\widehat{dF_j}\wedge\dots\wedge dF_p}\\
 & & \displaystyle{+\sum_{1\leq i<j\leq
p}(-1)^{i+j}F_0\,d\{F_i,\,F_j\}\wedge dF_1\wedge\dots\wedge
\widehat{dF_i}\wedge\dots\wedge \widehat{dF_j}\wedge\dots\wedge dF_p.} \\
 \end{array}$$ The Poisson homology space in degree $p$ is given by the formula
$$HP_p(A)=\textrm{Ker}\,\partial_p\,/\,\textrm{Im}\,\partial_{p+1}.$$
In particular, we have $HP_0(A)=A\,/\,\{A,\,A\}.$\\

\noindent $\bullet$ In the sequel, we will denote
$HP_0(\mathbb{C}[V]^W)$ by $HP_0(W)$ and $HH_0(\mathbb{C}[V]^W)$
by $HH_0(W)$.\\

\subsection{\textsf{Vectors of highest weight $0$}}

\noindent The aim of this section is to show that $S^W(2)$ is
isomorphic to $\go{sl}_2$ and that the vectors which do not belong
to $\{S^W,\,S^W\}$, are among the vectors of highest
weight $0$ of the $\go{sl}_2-$module $S^W$, an observation which will simplify the calculations.\\

\begin{Pro}\label{sl2}$\\$
For $B_n$ ($n\geq 2$) and $D_n$ ($n\geq 3$), the subspace $S^W(2)$
is isomorphic to $\go{sl}_2$. More precisely $S^W(2)=\langle E,\
F,\ H\rangle$ with the relations $\{H,\ E\}=2E,\ \{H,\ F\}=-2F$
and $\{E,\ F\}=H$, where the elements $E,\ F,\ H$ are explicitly
given by
$$\begin{array}{ccc}
  E=\frac{n}{2}R_n(x_1^2)=\frac{1}{2}\,\mathbf{x}\cdot \mathbf{x} & F=-\frac{n}{2}R_n(y_1^2)=-\frac{1}{2}\,\mathbf{y}\cdot \mathbf{y} &
  H=-n\,R_n(x_1y_1)=-\mathbf{x}\cdot \mathbf{y}. \\
\end{array}$$
For $D_2$, we have $S^W(2)=\langle E,\,F,\,H \rangle\oplus \langle
E',\,F',\,H' \rangle$ (direct sum of two Lie algebras isomorphic to $\go{sl}_2$).\\
So the spaces $S$ and $S^W$ are $\go{sl}_2-$modules.
\end{Pro}

\underline{Proof:}\\
We demonstrate the proposition for $B_n$, the proof being
analogous for $D_n$.\\
$\bullet$ As $x_1^2$ is invariant under sign changes, we may write
$R_n(x_1^2)=\frac{1}{n!}\sum_{\sigma\in\go{S}_n}\sigma\cdot
x_1^2$. Moreover, we have the partition $\go{S}_n=\coprod_{j=1}^n
A_j$, where $A_j:=\{\sigma\in\go{S}_n\ /\ \sigma(1)=j\}$ has
cardinality $(n-1)!$.\\
Thus $R_n(x_1^2)=\frac{(n-1)!}{n!}\sum_{j=1}^n x_j^2$. We proceed
likewise with $R_n(y_1^2)$ and $R_n(x_1y_1)$.\\
$\bullet$ We obviously have $S^W(2)\supset\langle E,\ F,\
H\rangle$.\\
Moreover, $R(x_j^2)=R(x_1^2)$, $R(y_j^2)=R(y_1^2)$ and
$R(x_jy_j)=R(x_1y_1)$. Last, if $i\neq j$, then $x_ix_j$, $y_iy_j$
and $x_iy_j$ are mapped to their opposite by the $i-$th sign
change $s_i$, and \begin{center} $W=s_i\cdot\langle s_1,\ \dots,\
s_{i-1},\ s_{i+1},\ \dots,\ s_n\rangle\cdot\go{S}_n\sqcup \langle
s_1,\ \dots,\ s_{i-1},\ s_{i+1},\ \dots,\
s_n\rangle\cdot\go{S}_n$,
\end{center}
thus $R_n(x_ix_j)=R_n(y_iy_j)=R_n(x_iy_j)=0$. Hence
$S^W(2)\subset\langle E,\ F,\ H\rangle$.\\
$\bullet$ We have $\nabla E=\left(\begin{array}{c}
  \mathbf{x} \\
  \mathbf{0} \\
\end{array}
\right),\ \nabla F=\left(\begin{array}{c}
  \mathbf{0} \\
  -\mathbf{y} \\
\end{array}
\right)$ and $\nabla H=\left(\begin{array}{c}
  -\mathbf{y} \\
  -\mathbf{x} \\
\end{array}
\right)$,\\so $\{E,\ F\}=H,\ \{H,\ E\}=2E$ and $\{H,\ F\}=-2F$.$\blacksquare$\\

\noindent We denote by $S_{\go{sl}_2}$ the set of vectors of
highest weight $0$, i.e.
the set of elements of $S$ which are annihilated by the action of $\go{sl}_2$.\\
For $j\in \mathbb{N}$, we denote by $S(j)_{\go{sl}_2}$
the elements of $S_{\go{sl}_2}$ of degree $j$.\\
As the action of $W$ and the action of $\go{sl}_2$ commute, we may
write
 $(S^W)_{\go{sl}_2}=(S_{\go{sl}_2})^W=S^W_{\go{sl}_2}$ and likewise for~$S^W(j)_{\go{sl}_2}$.\\

\begin{Pro}\label{deg0&4}$\\$
$\bullet$ If $S^W$ contains no element of degree $1$, then the
vectors of highest weight $0$ of degree $0$ do not belong to
$\{S^W,\ S^W\}$.\\
$\bullet$ Let $W$ be of type $B_n$ ($n\geq 2$) or $D_n$ ($n\geq
3$). If $S^W$ contains no element of degree $1,\ 3,\ $, then the
vectors of highest weight $0$ of degree $4$ do not belong to
$\{S^W,\ S^W\}$.
\end{Pro}

\underline{Proof:}\\
$\bullet$ The Poisson bracket being homogeneous of degree $-2$, we
have $S^W(0)\cap \{S^W,\ S^W\}=\{S^W(2),\
S^W(0)\}=\{0\}$.\\
$\bullet$ Similarly, $S^W(4)\cap \{S^W,\ S^W\}=\{S^W(0),\
S^W(6)\}+\{S^W(2),\ S^W(4)\}=\{\go{sl}_2,\ S^W(4)\}$, according to
Proposition \ref{sl2}. With the decomposition of the
$\go{sl}_2-$modules, we have \hbox{$S^W(4)=\bigoplus_{m\in
\mathbb{N}}V(m)$,} with $\{\go{sl}_2,\ V(m)\}=V(m)$ if $m\in
\mathbb{N}^*$ and $\{\go{sl}_2,\ V(0)\}=\{0\}$. So if \hbox{$a\in
S^W(4)\cap \{S^W,\ S^W\}$,} then $a\in \bigoplus_{m\in
\mathbb{N}^*}V(m)$.$\blacksquare$\\

\begin{Pro}\label{actionsl2}$\\$
For every monomial $M=\mathbf{x}^\mathbf{i}\mathbf{y}^\mathbf{j}$,
we have
\begin{center}
$\begin{array}{rcl}
  \{H,\ M\} & = & (|\mathbf{i}|-|\mathbf{j}|)M=(\emph{deg}_x(M)-\emph{deg}_y(M))M \\
  \{E,\ M\} & = & \sum_{k=1}^n j_k \,x_1^{i_1}\dots
x_{k-1}^{i_{k-1}}x_{k}^{i_{k}+1}x_{k+1}^{i_{k+1}}\dots
x_{n}^{i_{n}}y_1^{j_1}\dots
y_{k-1}^{j_{k-1}}y_{k}^{j_{k}-1}y_{k+1}^{j_{k+1}}\dots
y_{n}^{j_{n}}, \\
  \{F,\ M\} & = & \sum_{k=1}^n i_k \,x_1^{i_1}\dots
x_{k-1}^{i_{k-1}}x_{k}^{i_{k}-1}x_{k+1}^{i_{k+1}}\dots
x_{n}^{i_{n}}y_1^{j_1}\dots
y_{k-1}^{j_{k-1}}y_{k}^{j_{k}+1}y_{k+1}^{j_{k+1}}\dots
y_{n}^{j_{n}}. \\
\end{array}$
\end{center}
In particular, every vector of highest weight $0$ is of even
degree.
\end{Pro}

\underline{Proof:} This results from a simple calculation.\\

\begin{Rq}$\\$
Let $j\in \mathbb{N}$ and $P\in S^W(j)$. According to the
decomposition of $\go{sl}_2-$module $S^W(j)$ in weight subspaces,
we may write $P=\sum_{k=-m}^m P_k$ with $\{H,\,P_k\}=k\,P_k$. So
we have \fbox{$\frac{S^W(j)}{\{S^W,\ S^W\}\cap S^W(j)} \
=\frac{S^W(j)_{\go{sl}_2}}{\{S^W,\ S^W\}\cap
S^W(j)_{\go{sl}_2}}$}.\\
Thus the vectors which do not belong to $\{S^W,\ S^W\}$ are to be
found among the vectors of highest weight $0$.\\
\end{Rq}

\noindent The following property is a generalization of
Proposition 3 proved by J. Alev and L. Foissy in \cite{AF06}.
It enables us to know the Poincar\'e series of the algebra $S_{\go{sl}_2}$.\\

\begin{Pro}\label{alev}$\\$
For $l\in \mathbb{N}$, we have $S_{\go{sl}_2}(2l+1)=\{0\}$ and
$\dim\,S_{\go{sl}_2}(2l)=(C_{l+n-1}^{n-1})^2-C_{l+n}^{n-1}C_{l+n-2}^{n-1}$.\\
\end{Pro}

\noindent The following result is important for solving the
equation of Berest-Etingof-Ginzburg, because it gives a
description of the space of vectors of highest weight $0$, space
in which we will search the solutions of this equation. In the
proof of this proposition, we use the articles
\cite{DCP76} and \cite{GK04} concerning the pfaffian algebras.\\

\begin{Pro}\label{pfaff}$\\$
For $i\neq j$, set $X_{i,j}:=x_iy_j-y_ix_j$. Then the algebra
$\mathbb{C}[\mathbf{x},\ \mathbf{y}]_{\go{sl}_2}$ is the algebra
generated by the $X_{i,j}$'s for $(i,\ j)\in\llbr 1,\
n\rrbr^2$. We denote this algebra by $\mathbb{C}\langle X_{i,j}\rangle$.\\
This algebra is not a polynomial algebra for $n\geq 4$ (e. g.
$X_{1,2}X_{3,4}-X_{1,3}X_{2,4}+X_{2,3}X_{1,4}=0$).
\end{Pro}

\underline{Proof:}\\
$\bullet$ The inclusion $\mathbb{C}\langle
X_{i,j}\rangle\subset\mathbb{C}[\mathbf{x},\
\mathbf{y}]_{\go{sl}_2}$ being obvious, all that we have to do is
to show that the Poincar\'e series of both spaces are equal,
knowing that the one for $\mathbb{C}[\mathbf{x},\
\mathbf{y}]_{\go{sl}_2}$
is already given by Proposition \ref{alev}.\\
$\bullet$ Consider the vectors $u_j:=\left(
\begin{array}{c}
  x_j \\
  y_j \\
\end{array}
\right)$ for $j=1\dots n$, in the symplectic space $\mathbb{C}^2$
endowed with the standard symplectic form $\langle\cdot\rangle$
defined by the matrix $J:=\left(
\begin{array}{cc}
  0 & 1 \\
  -1 & 0 \\
\end{array}
\right)$. Let $\{T_{i,j}\ /\ 1\leq i<j\leq n\}$ be a set of
indeterminates, and let $\widetilde{T}$ be the antisymmetric
matrix the general term of which is $T_{i,j}$ if $i<j$. Then,
according to section 6 of \cite{DCP76}, the ideal $I_2$ of
relations between the $\langle u_i,\,u_j\rangle$ (i.e. between the
$X_{i,j}$) is generated by the pfaffian minors of $\widetilde{T}$
of size $4\times 4$. Set $PF:=\mathbb{C}\langle X_{i,j}\rangle$.
So we have $PF\simeq
\mathbb{C}[\left(T_{i,j}\right)_{i<j}]\,/\,I_2=:PF_0$, i.e. the
algebra $PF$ is isomorphic to the pfaffian algebra $PF_0$. Its
Poincar\'e series is given in section 4 of \cite{GK04} by
$$\dim\,PF_0(m)=(C_{m+n-2}^{m})^2-C_{m+n-2}^{m-1}C_{m+n-2}^{m+1}.$$
So we have
$\dim\,PF(2l)=(C_{l+n-2}^{l})^2-C_{l+n-2}^{l-1}C_{l+n-2}^{l+1}$.
We verify that
$$\dim\,PF(2l)=(C_{l+n-1}^{n-1})^2-C_{l+n}^{n-1}C_{l+n-2}^{n-1}=\dim\,\mathbb{C}[\mathbf{x},\
\mathbf{y}]_{\go{sl}_2}(2l).$$
We have obviously $\dim\,PF(2l+1)=0=\dim\,\mathbb{C}[\mathbf{x},\ \mathbf{y}]_{\go{sl}_2}(2l+1)$.
Hence the equality of the Poincar\'e series.$\blacksquare$\\

\subsection{\textsf{Equation of Berest-Etingof-Ginzburg}}

\noindent We study the functionnal equation introduced by Y.
Berest, P. Etingof and V. Ginzburg in \cite{BEG04}. The point is
that solving this equation, in the space $S^W_{\go{sl}_2}$, is
equivalent to the determination of the quotient
$\frac{S^W}{\{S^W,\ S^W\}}$, that is to say the
computation of the Poisson homology space in degree $0$ of $S^W$.\\

\begin{Lem}\label{etingof}(Berest - Etingof - Ginzburg)$\\$
We consider $\mathbb{C}^{2n}$, endowed with its standard symplectic form, denoted by $\langle\cdot,\,\cdot\rangle$. Let $j\in \mathbb{N}$.\\
Let $S:=\mathbb{C}[\mathbf{x},\mathbf{y}]=\mathbb{C}[\mathbf{z}]$,
and let $\mathcal{L}_j:=\left(\frac{S^W(j)}{\{S^W,\ S^W\}\cap
S^W(j)}\right)^*$ be the linear dual of
$\frac{S^W(j)}{\{S^W,\ S^W\}\cap S^W(j)}$.\\
Then $\mathcal{L}_j$ is isomorphic to the vector space of
polynomials $P\in \mathbb{C}[\mathbf{w}]^W(j)$ satisfying the
following equation:
\begin{equation}\label{equaetingof}
\forall\ \mathbf{w},\ \mathbf{w}'\in \mathbb{C}^{2n},\ \sum_{g\in
W}\langle \mathbf{w},\ g\mathbf{w}'\rangle\
P(\mathbf{w}+g\mathbf{w}')=0
\end{equation}
\end{Lem}

\underline{Proof:}\\
$\bullet$ For
$\mathbf{w}=(\mathbf{u},\,\mathbf{v})\in\mathbb{C}^{2n}$ and
$\mathbf{z}=(\mathbf{x},\,\mathbf{y})\in\mathbb{C}^{2n}$, we set
$L_\mathbf{w}(\mathbf{z}):=\sum_{g\in W}e^{\langle \mathbf{w},\
g\mathbf{z} \rangle}$.\\
So we have $\{L_\mathbf{w}(\mathbf{z}),\
L_\mathbf{w'}(\mathbf{z})\}=\nabla_\mathbf{x}L_\mathbf{w}(\mathbf{z})
\cdot\nabla_\mathbf{y}L_\mathbf{w'}(\mathbf{z})-\nabla_\mathbf{y}L_\mathbf{w}(\mathbf{z})
\cdot\nabla_\mathbf{x}L_\mathbf{w'}(\mathbf{z})$ $\\$

\noindent We deduce the formula
\begin{equation}\label{demet7}
\{L_\mathbf{w}(\mathbf{z}),\
L_\mathbf{w'}(\mathbf{z})\}=\sum_{g\in W}\langle \mathbf{w},\
g\mathbf{w'}\rangle L_{\mathbf{w}+g\mathbf{w'}}(\mathbf{z}).
\end{equation}

\noindent $\bullet$ Moreover, $L_\mathbf{w}(\mathbf{z})$ is a
power series in $\mathbf{w}$, the coefficients of which generate
$S^W$:\\
\begin{equation}\label{demet1}
L_\mathbf{w}(\mathbf{z})=\sum_{p=0}^\infty
\frac{|W|}{p!}\,R_n\left[\,\left(\sum_{i=1}^n
y_iu_i-x_iv_{i}\right)^p\,\right]
\end{equation}
The coefficients of the series are the images by $R_n$ of the
elements of the canonical basis of $S$.\\

\noindent Remark: in the case of $B_n$, there is no invariant of
odd degree, so we have $L_\mathbf{w}(\mathbf{z})=\sum_{g\in
W}\textmd{ch}(\langle
\mathbf{w},\ g\mathbf{z}\rangle)$.\\

\noindent$\triangleright$ Set
$M_p(\mathbf{z})=\{\mathbf{z}^\mathbf{i}\ /\ |\mathbf{i}|=p\}$ and
$M_p(\mathbf{w})=\{\mathbf{w}^\mathbf{i}\ /\
|\mathbf{i}|=p\}$.\\
For a monomial $m=\mathbf{x}^i\mathbf{y}^j\in M_p(\mathbf{z})$, let $\widetilde{m}=\mathbf{u}^j\mathbf{v}^i$.\\
Similarly, for a monomial $m=\mathbf{u}^i\mathbf{v}^j\in
M_p(\mathbf{w})$, let
$\overline{m}=\mathbf{x}^j\mathbf{y}^i$.\\
So, for $m\in M_p(\mathbf{z})$, we have
$\overline{\widetilde{m}}=m$,
and for $m\in M_p(\mathbf{w})$, we have $\widetilde{\overline{m}}=m$.\\

\noindent The series $L_\mathbf{w}(\mathbf{z})$ may then be
written as
\begin{equation}\label{demet2}
L_\mathbf{w}(\mathbf{z})=|W|+\sum_{j=1}^\infty\sum_{m_j\in
M_j(\mathbf{w})}\alpha_{m_j}R_n(\overline{m_j})m_j=|W|+\sum_{j=1}^\infty\sum_{m_j\in
M_j(\mathbf{z})}\alpha_{\widetilde{m_j}}R_n(m_j)\widetilde{m_j}=|W|+\sum_{j=1}^\infty
L_\mathbf{w}^j(\mathbf{z}),
\end{equation}
with $\alpha_{m_j}\in \mathbb{Q}^*$.\\
$\triangleright$ Now $\left(\sum_{i=1}^n
y_iu_i-x_iv_{i}\right)^p=\sum_{|\mathbf{a}|+|\mathbf{b}|=p}(-1)^{|\mathbf{b}|}
C_p^{\mathbf{a},\mathbf{b}}\mathbf{x}^\mathbf{b}\mathbf{y}^\mathbf{a}\mathbf{u}^\mathbf{a}\mathbf{v}^\mathbf{b},$
where $C_p^{\mathbf{a},\mathbf{b}}=\frac{p!}{a_1!\dots
a_n!b_1!\dots b_n!}$ is the multinomial coefficient, therefore
according to formula (\ref{demet1}), we have
\begin{equation}\label{demet3}
L_\mathbf{w}(\mathbf{z})=|W|+\sum_{p=1}^\infty
\,\sum_{|\mathbf{a}|+|\mathbf{b}|=p}(-1)^{|\mathbf{b}|}\frac{|W|}{p!}
C_p^{\mathbf{a},\mathbf{b}}R_n\left(\mathbf{x}^\mathbf{b}\mathbf{y}^\mathbf{a}\right)\mathbf{u}^\mathbf{a}\mathbf{v}^\mathbf{b}
\end{equation}
By collecting the formulae (\ref{demet2}) and (\ref{demet3}), we
obtain
\begin{equation}\label{demet4}
\alpha_{\mathbf{u}^\mathbf{a}\mathbf{v}^\mathbf{b}}=(-1)^{|\mathbf{b}|}\frac{|W|}{p!}C_p^{\mathbf{a},\mathbf{b}}.
\end{equation}
$\bullet$ We identify $\mathcal{L}_{j}$ with the vector space of
linear forms on $S^W(j)$ which vanish on $\{S^W,\
S^W\}\cap S^W(j)$.\\
Define the map
\begin{equation}\label{equaetingofd6}
\begin{array}{rcl}
  \pi\ :\ \mathcal{L}_{j} & \rightarrow & \displaystyle{\{P\in \mathbb{C}[\mathbf{w}]^W(j)\ /\ \forall\ \mathbf{w},\
\mathbf{w}'\in \mathbb{C}^{2n},\ \sum_{g\in W}\langle \mathbf{w},\
g\mathbf{w}'\rangle\ P(\mathbf{w}+g\mathbf{w}')=0\}} \\
  f & \mapsto & \pi_f:=f(L_\mathbf{w}^j). \\
\end{array}
\end{equation}
Then $\pi$ is well defined: indeed $L_\mathbf{w}^j$ is a
polynomial in $\mathbf{z}$ of degree $j$ with coefficients in
$\mathbb{C}[\mathbf{w}]$, and explicitly, we have
\begin{equation}\label{demet7bis}
f(L_\mathbf{w}^j)=\sum_{m_j\in
M_j(\mathbf{z})}\alpha_{\widetilde{m_j}}f(R_n(m_j))\widetilde{m_j}\in
\mathbb{C}[\mathbf{w}].
\end{equation}
$\triangleright$ If two monomials $m_j,\ m_j'\in M_j(\mathbf{z})$
belong to a same orbit under the action of $\go{S}_n$, then the
coefficients $\alpha_{\widetilde{m_j}}f(R_n(m_j))$ and
$\alpha_{\widetilde{m_j'}}f(R_n(m_j'))$ of $m_j$ and $m_j'$ are
the same, thus $f(L_\mathbf{w}^j)$ is invariant under $W$.\\
$\triangleright$ Besides, $\pi_f$ is solution of $E_n(P)=0$:
indeed, we may extend $f$ as a linear map defined on
$\frac{S^W}{\{S^W,\ S^W\}}=\bigoplus_{i=0}^\infty
\frac{S^W(i)}{\{S^W,\ S^W\}\cap S^W(i)}$, by setting $f=0$ on
$\frac{S^W(i)}{\{S^W,\ S^W\}\cap S^W(i)}$ for $i\neq
j$.\\
Then, according to (\ref{demet7}), we have the equality
\begin{center}
$0=f\left(\{L_\mathbf{w},\ L_\mathbf{w'}\}\right)=\sum_{g\in
W}\langle \mathbf{w},\ g\mathbf{w'}\rangle
f\left(L_{\mathbf{w}+g\mathbf{w'}}\right)$, \end{center} hence
$\sum_{g\in W}\langle \mathbf{w},\ g\mathbf{w'}\rangle
f\left(L_{\mathbf{w}+g\mathbf{w'}}^j\right)=0$. So, the polynomial
$f(L_\mathbf{w}^j)\in \mathbb{C}[\mathbf{w}]$
satisfies equation (\ref{equaetingof}).\\
$\bullet$ Define the map
\begin{equation}\label{equaetingofd6}
\begin{array}{rcl}
  \displaystyle{\varphi\ :\ \{P\in \mathbb{C}[\mathbf{w}]^W(j)\ /\ \forall\ \mathbf{w},\
\mathbf{w}'\in \mathbb{C}^{2n},\ \sum_{g\in W}\langle \mathbf{w},\
g\mathbf{w}'\rangle\ P(\mathbf{w}+g\mathbf{w}')=0\}} & \rightarrow & \mathcal{L}_{j} \\
  \displaystyle{P=\sum_{m_j\in M_j(\mathbf{w})}\beta_{m_j}m_j} & \mapsto & \left(\varphi_P\ :\ R_n(\overline{m_j})\mapsto \frac{\beta_{m_j}}{\alpha_{m_j}}\right). \\
\end{array}
\end{equation}
$\triangleright$ For $f\in \mathcal{L}_j$, we have
\begin{center}
$\varphi_{\pi_f}\left(R_n(\overline{m_j})\right)=\frac{\alpha_{m_j}f\left(R_n(\overline{m_j})\right)}{\alpha_{m_j}}=f\left(R_n(\overline{m_j})\right)$,
\end{center}
thus $\varphi_{\pi_f}=f$.\\
$\triangleright$ For $\displaystyle{P=\sum_{m_j\in
M_j(\mathbf{w})}\beta_{m_j}m_j\in \mathbb{C}[\mathbf{w}]^W(j)}$,
we have $\displaystyle{P=\sum_{m_j\in
M_j(\mathbf{z})}\beta_{\widetilde{m_j}}\widetilde{m_j}}$, so if
$m_j\in M_j(\mathbf{z})$, then
\mbox{$\varphi_P(R_n(m_j))=\frac{\beta_{\widetilde{m_j}}}{\alpha_{\widetilde{m_j}}}$}.
Consequently,
\begin{center}
$\pi_{\varphi_P}=\sum_{m_j\in
M_j(\mathbf{z})}\alpha_{\widetilde{m_j}}\varphi_P(R_n(m_j))\widetilde{m_j}
=\sum_{m_j\in
M_j(\mathbf{z})}\alpha_{\widetilde{m_j}}\frac{\beta_{\widetilde{m_j}}}{\alpha_{\widetilde{m_j}}}\widetilde{m_j}=P.$
\end{center}
So $\pi$ is bijective and its inverse is $\varphi$.\\
$\triangleright$ All we have to do is to show that $\varphi_P$
vanishes on
$\{S^W,\ S^W\}\cap S^W(j)$.\\
Let $P\in \mathbb{C}[\mathbf{w}]^W(j)$ be a solution of equation
(\ref{equaetingof}). Then as, $\pi_{\varphi_P}=P$, we have for
$k+l=j$,
\begin{center}
$0=\sum_{g\in W}\langle \mathbf{w},\ g\mathbf{w'}\rangle
P\left(\mathbf{w}+g\mathbf{w'}\right)=\varphi_P\left(\{L_\mathbf{w}^k,\
L_\mathbf{w'}^l\}\right)$
\end{center}
But
\begin{center}
$\{L_\mathbf{w}^k,\ L_\mathbf{w'}^l\}=\sum_{m_k\in
M_k(\mathbf{w})}\sum_{\mu_l\in
M_l(\mathbf{w'})}\alpha_{m_k}\alpha_{\mu_l}m_k\mu_l\,\{R_n(\overline{m_k}),\
R_n(\overline{\mu_l})\},$
\end{center}
so that
\begin{center}
$\sum_{m_k\in M_k(\mathbf{w})}\sum_{\mu_l\in
M_l(\mathbf{w'})}\alpha_{m_k}\alpha_{\mu_l}m_k\mu_l\,\varphi_P\left(\{R_n(\overline{m_k}),\
R_n(\overline{\mu_l})\}\right)=0.$
\end{center}
\vspace{.2cm}This last equality is equivalent to
\begin{center}
$\forall\ k+l=j,\ \varphi_P\left(\{R_n(\overline{m_k}),\
R_n(\overline{\mu_l})\}\right)=0,$
\end{center} which shows that $\varphi_P$ vanishes on $\{S^W,\ S^W\}\cap
S^W(j)$.$\blacksquare$\\

\vspace{1cm} \noindent The following corollary enables us to make
the equation of Berest-Etingof-Ginzburg more explicit.\\

\begin{Cor}\label{etingofbis}$\\$
We introduce $2n$ indeterminates, denoted by $z_1,\dots,\ z_n,\
t_1,\dots,\ t_n$, and we extend the Reynolds operator in a map
from $\mathbb{C}[\mathbf{x,\ y,\ z,\ t}]$ to itself which is
$\mathbb{C}[\mathbf{z,\ t}]-$linear. Then the vector space
$\mathcal{L}_j$ is isomorphic to the vector space of polynomials
$P\in S^W(j)$ satisfying the following equation:
\begin{equation}\label{equaetingofbis}
\begin{small}R_n \Big(\big(\sum_{i=1}^n z_iy_{i}-t_{i}x_i\big) \
P(z_1+x_1,\ \dots,\ z_n+x_n,\ t_{1}+y_{1},\ \dots,\
t_{n}+y_{n})\Big)=0
\end{small}
\end{equation}
i.e.
\begin{equation}\label{equaetingofter}
\fbox{$E_n(P):=R_n \Big((\mathbf{z}\cdot
\mathbf{y}-\mathbf{t}\cdot \mathbf{x}) \ P(\mathbf{x}+\mathbf{z},\
\mathbf{y}+\mathbf{t})\Big)=0$}
\end{equation}
\end{Cor}

\underline{Proof:}\\
\noindent We have $\langle \mathbf{w},\
\mathbf{w}'\rangle=\mathbf{w}\cdot(J\mathbf{w}')=\sum_{i=1}^n
(w_iw'_{n+i}-w_{n+i}w'_i)$. \noindent Then equation
(\ref{equaetingofbis}) is equivalent to
\begin{equation}\label{equaetingofd1}
\begin{array}{rl}
 \forall\ \mathbf{w},\ \mathbf{w}'\in \mathbb{C},
 & \sum_{g\in W}\sum_{i=1}^n (w_i\sum_{j=1}^n g_{ij}w'_{n+j}-w_{n+i}\sum_{j=1}^n g_{ij}w'_j) \\
   & \ P\Big(w_1+\sum_{j=1}^n g_{1j}w'_j,\ \dots,\ w_n+\sum_{j=1}^n
g_{nj}w'_j,\\
& w_{n+1}  +\sum_{j=1}^n g_{1j}w'_{n+j},\ \dots,\
w_{2n}+\sum_{j=1}^n g_{nj}w'_{n+j}\Big)=0 \\
\end{array}
\end{equation}

\noindent This means that the polynomial
\begin{small}
\begin{equation} \label{equaetingofd2}
\sum_{g\in W}\sum_{i=1}^n (z_i\sum_{j=1}^n
g_{ij}y_{j}-t_{i}\sum_{j=1}^n g_{ij}x_j) \ P\Big(z_1+\sum_{j=1}^n
g_{1j}x_j,\ \dots,\ z_n+\sum_{j=1}^n g_{nj}x_j,\
t_{1}+\sum_{j=1}^n g_{1j}y_{j},\ \dots,\ t_{n}+\sum_{j=1}^n
g_{nj}y_{j}\Big)
\end{equation}
\end{small}is zero.\\
\noindent This is equivalent to
\begin{small}
\begin{equation}\label{equaetingofd3}
\sum_{i=1}^n (z_i\,g\cdot y_{i}-t_{i}\,g\cdot x_i) \ \sum_{g\in
W}g\cdot \Big(P(z_1+x_1,\ \dots,\ z_n+x_n,\ t_{1}+y_{1},\ \dots,\
t_{n}+y_{n})\Big)=0,
\end{equation}
\end{small}
\noindent that is to say
\begin{small}
\begin{equation}\label{equaetingofd4}
R_n \Big(\big(\sum_{i=1}^n z_iy_{i}-t_{i}x_i\big) \ P(z_1+x_1,\
\dots,\ z_n+x_n,\ t_{1}+y_{1},\ \dots,\ t_{n}+y_{n})\Big)=0,
\end{equation}
\end{small}
\noindent where $R_n$ is the Reynolds operator extended in a $\mathbb{C}[\mathbf{z,\ t}]-$linear map.$\blacksquare$\\

\begin{Rq}\label{grindepnrq}$\\$
$\bullet$ Case of $B_n$: for a monomial $M\in \mathbb{C}[\mathbf{x},\ \mathbf{y}]$,\\
$\triangleright$ either there exists a sign change which sends
$M$ to its opposite, and then $R_n(M)=0$\\
$\triangleright$ or $M$ is invariant under every sign change and then $R_n(M)=\sum_{\sigma\in\go{S}_n}\sigma\cdot M$.\\
If $Q=R_n(P)$ with $P\in \mathbb{C}[\mathbf{x},\ \mathbf{y}]$,
then we may always assume that each monomial of $P$, in particular
$P$ itself, is invariant under the sign changes.\\
$\bullet$ Case of $D_n$: we have the same result, by considering
this time the sign changes of an even number of variables.\\
\end{Rq}

\noindent The aim of Proposition \ref{profond0} and its corollary
is to reduce drastically the space in which we search the
solutions of equation (\ref{equaetingofter}): indeed, instead of
searching the solutions in $S^W$, we may limit ourselves to the
space of the elements
 which are annihilated by the action of $\go{sl}_2$.\\

\begin{Pro}\label{profond0}$\\$
Let $P\in \mathbb{C}[\mathbf{x},\,\mathbf{y}]^W$. We consider the
element $E_n(P)$ defined by the formula (\ref{equaetingofter}) as
a polynomial in the indeterminates $\mathbf{z},\,\mathbf{t}$ and
with coefficients in
$\mathbb{C}[\mathbf{x},\,\mathbf{y}]$.\\
Then the coefficient of $z_1t_1$ in $E_n(P)$ is
$\frac{-1}{n}\,\{H,\,P\}$, that of $t_1^2$ is
$\frac{-1}{n}\{E,\,P\}$ and that of $z_1^2$
is~$\frac{1}{n}\{F,\,P\}$.
\end{Pro}

\underline{Proof:}\\
We carry out the proof for $B_n$. The method is the same for $D_n$.\\
$\bullet$ We denote by $c_{z_1t_1}(P)$ the coefficient of $z_1t_1$
in $E_n(P)$. Since the maps $P\mapsto c_{z_1t_1}(P)$ and $P\mapsto
\{H,\,P\}$ are linear, all we have to do is to prove the property
for $P$ of the form $P=R_n(M)$, where
$M=\mathbf{x}^\mathbf{i}\mathbf{y}^\mathbf{j}$ is a monomial which
we may assume invariant under the sign changes thanks to
remark \ref{grindepnrq}.\\
Then the formula (\ref{equaetingofter}) may be written
\begin{small}
$$\begin{array}{rcl}
  |W|\,E_n(M) & = & |W|\,R_n\left((\mathbf{z}\cdot \mathbf{y}-\mathbf{t}\cdot
\mathbf{x})(\mathbf{x}+\mathbf{z})^\mathbf{i}(\mathbf{y}+\mathbf{t})^\mathbf{j}\right) \\
    & = & \displaystyle{\sum_{c \in (\pm 1)^n}\sum_{\sigma\in\go{S}_n}c\cdot\left[(z_1y_{\sigma^{-1}(1)}+\dots+z_ny_{\sigma^{-1}(n)})
    \prod_{k=1}^n(z_k+x_{\sigma^{-1}(k)})^{i_k}(t_k+y_{\sigma^{-1}(k)})^{j_k}\right]} \\
     & & \displaystyle{- \sum_{c \in (\pm 1)^n}\sum_{\sigma\in\go{S}_n}c\cdot\left[(t_1x_{\sigma^{-1}(1)}+\dots+t_nx_{\sigma^{-1}(n)})
    \prod_{k=1}^n(z_k+x_{\sigma^{-1}(k)})^{i_k}(t_k+y_{\sigma^{-1}(k)})^{j_k}\right]} \\
\end{array}$$
\end{small}
$\bullet$ So the coefficient of $z_1t_1$ is given by
$$\begin{array}{rcl}
  |W|\,c_{z_1t_1}(M) & = & \displaystyle{\sum_{c \in (\pm
1)^n}\sum_{\sigma\in\go{S}_n}c\cdot\Big[y_{\sigma^{-1}(1)}\,\left(\prod_{k=1}^n
x_{\sigma^{-1}(k)}^{i_k}\right)\,j_1\,y_{\sigma^{-1}(1)}^{j_1-1}\left(\prod_{k=2}^n
y_{\sigma^{-1}(k)}^{j_k}\right)}\\
 & &  \ \ \ \ \ \ \ \ \ \ \ \ \ \ \ \displaystyle{-
x_{\sigma^{-1}(1)}\,i_1\,x_{\sigma^{-1}(1)}^{i_1-1}\,\left(\prod_{k=2}^n
x_{\sigma^{-1}(k)}^{i_k}\right)\left(\prod_{k=1}^n
y_{\sigma^{-1}(k)}^{j_k}\right)\Big]}\\
   & = & \displaystyle{|(\pm 1)^n|\,\sum_{\sigma\in\go{S}_n} (j_1-i_1)\left(\prod_{k=1}^n
x_{\sigma^{-1}(k)}^{i_k}y_{\sigma^{-1}(k)}^{j_k}\right)}\\
 & = & |W|\,(j_1-i_1)\,R_n(M).\\
\end{array}$$
Since $P=\frac{1}{n!}\,\sum_{\sigma\in\go{S}_n}\prod_{k=1}^n
x_k^{i_{\sigma(k)}}y_k^{j_{\sigma(k)}}$, we deduce that
$$\begin{array}{rcl}
  n!\ c_{z_1t_1}(P) & = & \displaystyle{\sum_{\sigma\in\go{S}_n}c_{z_1t_1}\left(\prod_{k=1}^n
x_k^{i_{\sigma(k)}}y_k^{j_{\sigma(k)}}\right)=\sum_{\sigma\in\go{S}_n}\,(j_{\sigma(1)}-i_{\sigma(1)})\,R_n\left(\prod_{k=1}^n
x_k^{i_{\sigma(k)}}y_k^{j_{\sigma(k)}}\right)} \\
  & = & \displaystyle{\sum_{\sigma\in\go{S}_n}\,(j_{\sigma(1)}-i_{\sigma(1)})\,R_n\left(M\right)=
  (n-1)!\,\sum_{k=1}^n\,(j_k-i_k)\,R_n\left(M \right)}\\
   & = & (n-1)!\,(\textrm{deg}_y(M)-\textrm{deg}_x(M))\,R_n\left(M \right)=-(n-1)!\,\{H,\,P\}. \\
\end{array}$$
$\bullet$ We proceed as for $z_1t_1$, by denoting by
$c_{t_1^2}(P)$ the coefficient of $t_1^2$ in $E_n(P)$. Then we
have
$$\begin{array}{rcl}
  |W|\,c_{t_1^2}(M) & = & \displaystyle{-\sum_{c \in (\pm
1)^n}\sum_{\sigma\in\go{S}_n}c\cdot\Big[x_{\sigma^{-1}(1)}\,\left(\prod_{k=1}^n
x_{\sigma^{-1}(k)}^{i_k}\right)\,j_1\,y_{\sigma^{-1}(1)}^{j_1-1}\,\left(\prod_{k=2}^n
y_{\sigma^{-1}(k)}^{j_k}\right)\Big]}\\
   & = & \displaystyle{-|(\pm 1)^n|\,j_1\,\sum_{\sigma\in\go{S}_n}x_{\sigma^{-1}(1)}^{i_1+1}y_{\sigma^{-1}(1)}^{j_1-1}\left(\prod_{k=2}^n
x_{\sigma^{-1}(k)}^{i_k}y_{\sigma^{-1}(k)}^{j_k}\right)}\\
 & = & \displaystyle{-|W|\,j_1\ R_n\left(x_1^{i_1+1}y_1^{j_1-1}\left(\prod_{k=2}^n
x_k^{i_k}y_k^{j_k}\right)\right)}.\\
\end{array}$$
Thus $$\begin{array}{rcl}
  n!\ c_{t_1^2}(P) & = & \displaystyle{\sum_{\sigma\in\go{S}_n}c_{t_1^2}\left(\prod_{k=1}^n
x_k^{i_{\sigma(k)}}y_k^{j_{\sigma(k)}}\right)=-\sum_{\sigma\in\go{S}_n}\,j_{\sigma(1)}\,R_n\left(x_1^{i_{\sigma(1)}+1}y_1^{j_{\sigma(1)}-1}\left(\prod_{k=2}^n
x_k^{i_{\sigma(k)}}y_k^{j_{\sigma(k)}}\right)\right)} \\
  & = & \displaystyle{-\sum_{p=1}^n\sum_{\substack{\sigma\in\go{S}_n \\
    \sigma(1)=p}}\,j_{\sigma(1)}\,R_n\left(x_1^{i_{\sigma(1)}+1}y_1^{j_{\sigma(1)}-1}\left(\prod_{k=2}^n
x_k^{i_{\sigma(k)}}y_k^{j_{\sigma(k)}}\right)\right)}\\
   & = & \displaystyle{-(n-1)!\,\sum_{p=1}^n\,j_p\,R_n\left(x_1^{i_1}\dots x_p^{i_p+1}\dots
   x_n^{i_n}y_1^{j_1}\dots y_p^{j_p-1}\dots y_n^{j_n}\right)}. \\
\end{array}$$
But $$\begin{array}{rcl}
  n!\ \{E,\,P\} & = & \displaystyle{\sum_{\sigma\in\go{S}_n}\{E,\,x_1^{i_{\sigma(1)}}\dots x_n^{i_{\sigma(n)}}
  y_1^{j_{\sigma(1)}}\dots y_n^{j_{\sigma(n)}}\}} \\
    & = & \displaystyle{\sum_{\sigma\in\go{S}_n} \sum_{p=1}^n j_{\sigma(p)}x_1^{i_{\sigma(1)}}\dots x_p^{i_{\sigma(p)}+1}\dots
   x_n^{i_{\sigma(n)}}y_1^{j_{\sigma(1)}}\dots y_p^{j_{\sigma(p)}-1}\dots y_n^{j_{\sigma(n)}}}  \\
 & = & \displaystyle{\sum_{p=1}^n \sum_{q=1}^n \sum_{\substack{\sigma\in\go{S}_n\\\sigma(p)=q}}  j_{\sigma(p)}x_1^{i_{\sigma(1)}}\dots x_p^{i_{\sigma(p)}+1}\dots
   x_n^{i_{\sigma(n)}}y_1^{j_{\sigma(1)}}\dots y_p^{j_{\sigma(p)}-1}\dots y_n^{j_{\sigma(n)}}}  \\
& = & \displaystyle{\sum_{q=1}^n j_q\,\sum_{p=1}^n
\sum_{\substack{\sigma\in\go{S}_n\\\sigma(p)=q}}
x_1^{i_{\sigma(1)}}\dots x_p^{i_{\sigma(p)}+1}\dots
   x_n^{i_{\sigma(n)}}y_1^{j_{\sigma(1)}}\dots y_p^{j_{\sigma(p)}-1}\dots y_n^{j_{\sigma(n)}}}  \\
& = & \displaystyle{n!\,\sum_{q=1}^n j_q\,R_n\left(x_1^{i_1}\dots
x_q^{i_q+1}\dots x_n^{i_n}y_1^{j_1}\dots y_q^{j_q-1}\dots y_n^{j_n}\right)}.\\
\end{array}$$
So $c_{t_1^2}(P)=\frac{-1}{n}\{E,\,P\}$. Similarly we show that $c_{z_1^2}(P)=\frac{1}{n}\{F,\,P\}$.$\blacksquare$\\

\begin{Cor}\label{profond}$\\$
$\bullet$ Let $P\in \mathbb{C}[\mathbf{x},\,\mathbf{y}]^W$. If $P$
satisfies equation (\ref{equaetingofter}), then $P$ is annihilated
by
$\go{sl}_2$, i.e. $P\in S^W_{\go{sl}_2}$.\\
$\bullet$ Therefore the vector space $\mathcal{L}_j$ is isomorphic
to the vector space of the polynomials $P\in S^W_{\go{sl}_2}(j)$
satisfying equation~(\ref{equaetingofter}).\\
Thus the determination of $\frac{S^W}{\{S^W,\ S^W\}} \
\left(=\frac{S^W_{\go{sl}_2}}{\{S^W,\ S^W\}\cap
S^W_{\go{sl}_2}}\right)$ is equivalent to the resolution, in
$S^W_{\go{sl}_2}$, of equation~(\ref{equaetingofter}).
\end{Cor}

\underline{Proof:}\\
Let $P\in \mathbb{C}[\mathbf{x},\,\mathbf{y}]^W$ satisfying
equation (\ref{equaetingofter}). Then all the coefficients of the
polynomial $E_n(P)\in
(\mathbb{C}[\mathbf{x},\,\mathbf{y}])[\mathbf{z},\,\mathbf{t}]$
are zero. In particular, according to Proposition \ref{profond0},
we have $\{H,\,P\}=\{E,\,P\}=\{F,\,P\}=0$. Hence $P\in S^W_{\go{sl}_2}$.\\
The second point results from Corollary \ref{etingofbis} and from the first point.$\blacksquare$\\

\noindent We end this section by defining two variants of the
equation of Berest-Etingof-Ginzburg: these are technical tools
which enable us to eliminate some variables and thus to solve
equation (\ref{equaetingofter}) more easily.\\

\begin{Def}$\\$
We define the (intermediate) map
\begin{center}
$\begin{array}{rcl}
  s_{int}^n\ :\ \mathbb{C}[\mathbf{x,\ y,\ z,\ t}] & \rightarrow & \mathbb{C}[\mathbf{x,\ y,\ z,\ t}] \\
  P & \mapsto & P(\mathbf{0}\;\ \mathbf{y}\;\ \mathbf{z}\;\ t_1,\ \mathbf{0}), \\
\end{array}$
\end{center}
and we set
\begin{equation}\label{equaetingofsubsint}
E_{int}^n(P):=s_{int}^n(E_n(P)).
\end{equation}
\noindent Similarly, we define the map
\begin{center}
$\begin{array}{rcl}
  s_n\ :\ \mathbb{C}[\mathbf{x,\ y,\ z,\ t}] & \rightarrow & \mathbb{C}[\mathbf{x,\ y,\ z,\ t}] \\
  P & \mapsto & P(\mathbf{0}\;\ y_1,\ \mathbf{0}\;\ \mathbf{z}\;\ t_1,\ \mathbf{0}), \\
\end{array}$
\end{center}
and we set
\begin{equation}\label{equaetingofsubs}
E'_n(P):=s_n(E_n(P)).
\end{equation}
This last equation is equation (\ref{equaetingofter}) after the
substitution
\begin{center}
$x_1=\dots=x_n=y_2=\dots=y_n=t_2=\dots=t_n=0$.
\end{center}$\\$
\end{Def}

\begin{Rq}\label{rqev}$\\$
If $P$ satisfies equation (\ref{equaetingofter}), it satisfies
obviously equation (\ref{equaetingofsubs}).\\
\end{Rq}

\noindent In the case where $n$ is an odd integer, the vectors of
highest weight $0$ of even degree are the same for $B_n$ and
$D_n$, and equations (\ref{equaetingofsubs}) are identical for
both types. Moreover, the link between equations
(\ref{equaetingofter}) for $B_n$ and $D_n$ enables us to prove the
inequality $\dim\,HP_0(D_{n})\leq
\dim\,HP_0(B_{n})$. It is the purpose of the two following propositions.\\

\begin{Pro}\label{lienBnDn}$\\$
By abuse of notation, we denote by $S^{B_n}(2p)$ (resp.
$S^{D_n}(2p)$) the set of invariant elements of degree $2p$ in
type $B_n$ (resp. $D_n$). Then we have
$S^{B_{2n+1}}(2p)=S^{D_{2n+1}}(2p)$,
$S^{B_{2n+1}}_{\go{sl}_2}(p)=S^{D_{2n+1}}_{\go{sl}_2}(p)$, and
equations (\ref{equaetingofsubs}) in
$S^{B_{2n+1}}_{\go{sl}_2}=S^{D_{2n+1}}_{\go{sl}_2}$ associated to both types are the same.\\
This result is false for the even indices: counter-example: $\dim\
S^{D_4}_{\go{sl}_2}(6)=1$ whereas $S^{B_4}_{\go{sl}_2}(6)=\{0\}$.
\end{Pro}

\underline{Proof:}\\
$\bullet$ We set $$\Phi_{2n+1}(P)=\sum_{\sigma\in
\go{S}_{2n+1}}\sigma\cdot P,\ \ \ \ \ \Psi_{2n+1}^B(P)=\sum_{g\in
(\pm 1)^{2n+1}}g\cdot P,\ \ \ \ \ \Psi_{2n+1}^D(P)=\sum_{g\in (\pm
1)^{2n}}g\cdot P,$$ so that
$$R_{2n+1}^B(P)=\frac{1}{|B_{2n+1}|}\Phi_{2n+1}\circ\Psi_{2n+1}^B,\
\ \textrm{and}\ \
R_{2n+1}^D(P)=\frac{1}{|D_{2n+1}|}\Phi_{2n+1}\circ\Psi_{2n+1}^D.$$
We obviously have
$\Psi_{2n+1}^B(S(2p))\subset\Psi_{2n+1}^D(S(2p))$.\\
Conversely, since $\Psi_{2n+1}^D(S(2p))$ is spanned by the
elements of the form $\Psi_{2n+1}^D(\mathbf{m})$ with
$\mathbf{m}\in S(2p)$ monomial, all we have to do is to show that
$\Psi_{2n+1}^D(\mathbf{m})$ belongs to $\Psi_{2n+1}^B(S(2p))$,
i.e. $\Psi_{2n+1}^D(\mathbf{m})$ is invariant under the sign
changes. Now $\mathbf{m}=x_1^{i_1}\dots
x_{2n+1}^{i_{2n+1}}y_1^{j_1}\dots y_{2n+1}^{j_{2n+1}}$ with
$\sum_{k=1}^{2n+1}(i_k+j_k)=2p$, therefore at least one of the
$i_k+j_k$
is even. Let's denote by $l$ the corresponding index.\\
So, for every $k\neq l$, we have
$s_k(\mathbf{m})=(-1)^{i_k+j_k}\mathbf{m}=s_{k,l}(\mathbf{m})$ and
$s_l(\mathbf{m})=\mathbf{m}$. But
$$\Psi_{2n+1}^D(\mathbf{m})=\underbrace{\left(\sum_{q_1=0,1\dots
q_{2n+1}=0,1}(-1)^{q_1[(i_1+j_1)+(i_2+j_2)]+q_2[(i_2+j_2)+(i_3+j_3)]+\dots
+q_{2n}[(i_{2n}+j_{2n})+(i_{2n+1}+j_{2n+1})]}\right)}_{a_{\mathbf{m}}}\mathbf{m},$$
therefore
$s_k\left(\Psi_{2n+1}^D(\mathbf{m})\right)=\left\{\begin{array}{ll}
  a_{\mathbf{m}}s_{k,l}(\mathbf{m})=s_{k,l}(a_{\mathbf{m}}\mathbf{m})=s_{k,l}\left(\Psi_{2n+1}^D(\mathbf{m})\right) & \textrm{si}\ k\neq l \\
  a_{\mathbf{m}}\mathbf{m} & \textrm{si}\ k=l \\
\end{array}\right\}=\Psi_{2n+1}^D(\mathbf{m})$.\\
$\bullet$ So we have
$S^{D_n}(2p)=\Phi_{2n+1}\left(\Psi_{2n+1}^D(S(2p))\right)=\Phi_{2n+1}\left(\Psi_{2n+1}^B(S(2p))\right)=S^{B_n}(2p)$.\\
Hence $S^{B_n}_{\go{sl}_2}(2p)=S^{D_n}_{\go{sl}_2}(2p)$. Besides,
according to Proposition \ref{actionsl2},
$S^{B_n}_{\go{sl}_2}(2p+1)=S^{D_n}_{\go{sl}_2}(2p+1)=\{0\}$.\\
$\bullet$ For $P\in S^{B_n}_{\go{sl}_2}$, equation
(\ref{equaetingofsubs}) may be written
$$\sum_{i=1}^{2n+1}z_i\sum_{\substack{\sigma\in
\go{S}_n\\\sigma(1)=i}}\left[y_1\,P(z_1,\dots,\,z_{2n+1},\,t_1,\,0,\dots,\,0,\,\underbrace{y_1}_i,\,0,\dots,\,0)-y_1\,P(z_1,\dots,\,z_{2n+1},\,t_1,\,0,\dots,\,0,\,\underbrace{-y_1}_i,\,0,\dots,\,0)\right]=0.$$
It is equation (\ref{equaetingofsubs}) for $P\in
S^{D_n}_{\go{sl}_2}$.$\blacksquare$\\

\begin{Pro}\label{lienBnDnbis}$\\$
Let $P$ be invariant by sign changes. If $P$ is solution of
equation (\ref{equaetingofter}) for $D_n$, then
$P$ is solution of equation (\ref{equaetingofter}) for $B_n$.\\
In particular, we have $\dim\,HP_0(D_{2n+1})\leq
\dim\,HP_0(B_{2n+1})$.
\end{Pro}

\underline{Proof:}\\
Let $SB_n$ (resp. $SD_n$) be the group of sign changes of $B_n$
(resp. $D_n$). We may write $SB_n=SD_n\,\sqcup\,
SD_n\cdot s_1$.\\
Let $P$ be invariant by sign changes. So we have
$P=R_n^B(P)=R_n^D(P)$, and equation (\ref{equaetingofter}) for
$B_n$ (resp. $D_n$) may be written $E_n^B(P)=R_n^B(Q)$ (resp.
$E_n^D(P)=R_n^D(Q)$), with $Q=(\mathbf{z}\cdot
\mathbf{y}-\mathbf{t}\cdot
\mathbf{x})\,P(\mathbf{x}+\mathbf{z}\,\mathbf{y}+\mathbf{t})$.\\
If $P$ is solution of equation (\ref{equaetingofter}) for $D_n$,
then we have: $$\begin{array}{rcl}
  R_n^B(Q) & = & \displaystyle{\sum_{h\in SB_n}\sum_{\sigma\in \go{S}_n}(\sigma h)\cdot Q=\sum_{h\in SB_n}h\cdot\left(\sum_{\sigma\in \go{S}_n}\sigma\cdot Q\right)} \\
    & = & \displaystyle{\sum_{g\in SD_n}g\cdot\left(\sum_{\sigma\in \go{S}_n}\sigma\cdot Q\right)+\sum_{g\in SD_n}(gs_1)\cdot\left(\sum_{\sigma\in \go{S}_n}\sigma\cdot Q\right)} \\
    & = & \displaystyle{\sum_{g\in SD_n}g\cdot\left(\sum_{\sigma\in \go{S}_n}\sigma\cdot Q\right)+s_1\cdot\left[\sum_{g\in SD_n}g\cdot\left(\sum_{\sigma\in \go{S}_n}\sigma\cdot Q\right)\right]} \\
    & = & \displaystyle{R_n^D(Q)+s_1\cdot R_n^D(Q)=0.}\\
\end{array}$$
So, $P$ is solution of equation (\ref{equaetingofter}) for
$B_n$.\\
We deduce the claimed inequality, knowing that, according to
Proposition \ref{lienBnDn},
 $S^{B_{2n+1}}_{\go{sl}_2}=S^{D_{2n+1}}_{\go{sl}_2}$. $\blacksquare$\\

\subsection{\textsf{Construction of graphs attached to the  invariant polynomials}}

\noindent Let us recall the equality of Proposition \ref{pfaff}:
$S^W_{\go{sl}_2}=R_n\left(\mathbb{C}\langle
X_{i,j}\rangle\right)$. Moreover, according to Corollary
\ref{profond}, the computation of $HP_0(S^W)$ can be reduces to
solving Equation (\ref{equaetingofter}) in the space $S^W_{\go{sl}_2}$.\\
In order to have shorter and more visual notations, we represent
the polynomials of this space by graphs, by the method explained in definition \ref{defgra}.\\
But before, let us quote, for the particular case that we are
interested in, the fundamental result established by J. Alev, M.
A. Farinati, T. Lambre and A. L. Solotar in \cite{AFLS00}:\\

\begin{The}(Alev-Farinati-Lambre-Solotar)$\\$
For $k=0\dots 2n$, the dimension of $HH_k(A_n(\mathbb{C})^W)$ is
the number of conjugacy classes of $W$ admitting the eigenvalue
$1$ with the multiplicity $k$.
\end{The}

\noindent By specializing to the cases of $B_n$ and $D_n$, we
obtain:

\begin{Cor}(Alev-Farinati-Lambre-Solotar)$\\$
$\bullet$ For type $B_n$, the dimension of
$HH_0(A_n(\mathbb{C})^W)$ is
the number of partitions $\pi(n)$ of the integer $n$.\\
$\bullet$ For type $D_n$, the dimension of
$HH_0(A_n(\mathbb{C})^W)$ is the number of partitions
$\widetilde{\pi}(n)$ of the integer $n$ having an even number of parts.\\
\end{Cor}

\noindent The conjecture of J. Alev may be set forth as follows:\\

\begin{Conj}(Alev)$\\$
$\bullet$ For the type $B_n$, the dimension of $HP_0(S^W)$ equals
the number
 of partitions $\pi(n)$ of the integer $n$.\\
$\bullet$ For the type $D_n$, the dimension of $HP_0(S^W)$ equals
the
number of partitions $\widetilde{\pi}(n)$ of the integer $n$ having an even number of parts.\\
\end{Conj}

\noindent Now, let us show how to construct $\pi(n)$ solutions of
equation~(\ref{equaetingofter}) for the case of
$B_n$.\\

\begin{Def}\label{defgra} $\\$
For $i\neq j$, we note $X_{i,j}=x_iy_j-y_ix_j$.\\
To each element of the form
$M:=\prod_{i=1}^{n-1}\prod_{j=i+1}^{n}X_{i,j}^{2a_{i,j}}$,
we associate the (non-oriented) graph $\widetilde{\Gamma_M}$ such that\\
$\triangleright$ the set of vertices of $\widetilde{\Gamma_M}$ is
the set of indices $\{k\in\llbr 1,\,n\rrbr\ /\
\exists\,i\in \llbr 1,\,n\rrbr\ /\ a_{i,k}\neq 0\ \emph{or}\ a_{k,i}\neq 0\}$,\\
$\triangleright$ two vertices $i,\ j$ of $\widetilde{\Gamma_M}$
are connected by the edge $\xymatrix{i \ar@1{-}[r]^{a_{i,j}} & j}$
if $a_{i,j}\neq 0$.
\end{Def}

\noindent $\bullet$ If $\sigma\in\go{S}_n$, then the graph
$\widetilde{\Gamma_{\sigma\cdot M}}$ is obtained by permuting the
vertices of
$\widetilde{\Gamma_{M}}$.\\
So, by replacing each vertex by the symbol $\bullet$, we obtain a
graph $\Gamma_M$ such that the map $M\mapsto \Gamma_M$ is constant
on every orbit under the action of $B_n$ (resp. $D_n$). So we may
associate this graph to the element $R_n
\left(\prod_{i=1}^{n-1}\prod_{j=i+1}^{n}X_{i,j}^{2a_{i,j}}\right)$.\\
To a linear combination $\sum_{k=1}^p \alpha_k M_k$, we associate the graph $\sum_{k=1}^p \alpha_k\, \Gamma_{M_k}$.\\

\noindent $\bullet$ We may extend this definition to elements of
the form $M:=\prod_{i=1}^{n-1}\prod_{j=i+1}^{n}X_{i,j}^{b_{i,j}}$
by denoting an edge by $\xymatrix{\bullet
\ar@1{-}[r]^{\frac{b_{i,j}}{2}} & \bullet}$ if $b_{i,j}$ is even
and by $\xymatrix{\bullet \ar@1{..}[r]^{\frac{b_{i,j}+1}{2}} &
\bullet}$ if $b_{i,j}$ is odd. Be careful ! We have for example
$\xymatrix{\bullet
\ar@1{..}[r] & \bullet}=0$.\\

\begin{Ex}$\\$
The polynomial $R_4(X_{1,2}^4X_{1,3}^2X_{1,4}^2)$ is represented
by the graph $
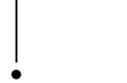
\xymatrix{\bullet \ar@2{-}[r] & \bullet \ar@1{-}[r]
\ar@1{-}[d] & \bullet \cr & \bullet & }$.\\
\end{Ex}

\noindent $\bullet$ If a graph contains only even edges (i.e. of
the form $\xymatrix{\bullet \ar@1{-}[r]^{a_{i,j}} & \bullet}$),
then
it is represented in $B_n$ and in $D_n$ by the same element.\\
This result is not valid in the case of graphs which contain odd
edges: for example, the element $
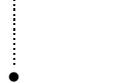
\xymatrix{\bullet \ar@1{..}[r] &
\bullet \ar@1{..}[r] \ar@1{..}[d] & \bullet \cr &
\bullet & }$ is zero in $B_4$, but different from zero in $D_4$.\\

\begin{Rq}$\\$
The graphs corresponding to polynomials obtained by operating the
Reynolds operator  for different indices on the same elements of
the algebra generated by the $X_{i,j}$'s are the same.\\
For example, the elements $R_3(X_{1,2}^2)$ and $R_{44}(X_{1,2}^2)$
are represented in this way by the same graph $\xymatrix{\bullet
\ar@1{-}[r] & \bullet}$. Propositions \ref{grindepn} and
\ref{grindepn2} show that this has no effect on the study of
equation (\ref{equaetingofter}) for $B_n$.\\
\end{Rq}

\begin{Pro}\label{grapheslin\'eaires}$\\$
For every $n\in \mathbb{N}^*$, the number of linear graphs without
loops and without isolated vertices is equal to the number of
partitions of $n$. (A multiple edge is viewed as a loop).
\end{Pro}

\underline{Proof:} immediate. To each partition $p$ of the form
$n=1p_1+2p_2+3p_3+\dots +np_n$, we associate the graph having
$p_j$ linear connected components with $j$ vertices.\\

\begin{Pro}\label{grindepn}$\\$
$\bullet$ Let $P\in \mathbb{C}[x_1,\dots,\ x_n,\
y_1,\dots,\ y_n]$. If $R_n(P)\neq 0$ then $R_{n+1}(P)\neq 0$.\\
$\bullet$ Let $P_1,\ \dots,\ P_m\in \mathbb{C}[x_1,\dots,\ x_n,\
y_1,\dots,\ y_n]$. If $R_n(P_1),\ \dots,\ R_n(P_m)$ are linearly
independent, then $R_{n+1}(P_1),\ \dots,\ R_{n+1}(P_m)$ are
linearly independent.
\end{Pro}

\underline{Proof:}\\
$\bullet$ We carry out the proof for $B_n$. We proceed likewise for $D_n$.\\
Let $P\in \mathbb{C}[x_1,\dots,\ x_n,\ y_1,\dots,\ y_n]$ such that
$R_n(P)\neq 0$. According to remark \ref{grindepnrq}, we may
assume that the terms of $P$ are invariant under sign changes.\\
We consider the set $\mathcal{T}_n$ of the terms of $R_n(P)$ that
we partition into orbits under the action of $\go{S}_n$: so we
have the equality $\mathcal{T}_n=\coprod_{j=1}^r\mathcal{O}_j$.
Consequently, $R_n(P)$ may be written $$R_n(P)=\sum_{j=1}^r
\alpha_j
R_n(M_j),$$ where $M_j\in \mathcal{O}_j$.\\
Let $c_{n+1}:=(1,\ \dots,\ n+1)\in\go{S}_{n+1}$, and $s_{n+1}$ the
$(n+1)-$th sign change, so that $B_{n+1}=\langle s_{n+1},\
c_{n+1}\rangle\cdot B_n$. Again by the invariance under sign
changes, we deduce
$$R_{n+1}(P)=\frac{1}{n+1}\sum_{j=1}^r \alpha_j \underbrace{\left(\sum_{k=0}^n
c_{n+1}^k\cdot R_n(M_j)\right)}_{t_j}.$$ Now if $i\neq j$, then
$t_i$ and $t_j$ belong to two distinct orbits under the action of
$\go{S}_{n+1}$, Therefore the $t_j$'s are linearly independent. So $R_{n+1}(P)\neq 0$.\\
$\bullet$ Let be $P_1,\ \dots,\ P_m\in \mathbb{C}[x_1,\dots,\
x_n,\ y_1,\dots,\ y_n]$ such that $R_n(P_1),\ \dots,\ R_n(P_m)$
are linearly independent. Let's consider a zero linear combination
$\sum_{j=1}^m \lambda_j R_{n+1}(P_j)=0$, i.e.
$R_{n+1}\left(\sum_{j=1}^m \lambda_j P_j\right)=0$. Then,
according to the first point, we have $\sum_{j=1}^m \lambda_j
R_n(P_j)=R_n\left(\sum_{j=1}^m \lambda_j P_j\right)=0$, so by
hypothesis, $\forall\ j\in\llbr1,\ m\rrbr,\
\lambda_j=0$.$\blacksquare$\\

\noindent The following proposition shows  the fact that, for a
graph, being solution of equation (\ref{equaetingofter}) for $B_n$
is independent of $n$, provided that $n$ is not smaller than the
number of vertices of the graph ! This proposition
justifies the $n$-independent notation of graphs.\\

\begin{Pro}\label{grindepn2}$\\$
Let $P\in \mathbb{C}[x_1,\dots,\ x_n,\ y_1,\dots,\ y_n]$.\\
$\bullet$ Case of $B_n$: if $R_n(P)$ satisfies the equation
$E_n\left(R_n(P)\right)=0$, then $R_{n+1}(P)\in
\mathbb{C}[x_1,\dots,\ x_{n+1},\ y_1,\dots,\ y_{n+1}]$ satisfies
the equation $E_{n+1}\left(R_{n+1}(P)\right)=0$.\\
$\bullet$ Case of $D_n$: if $R_n(P)$ satisfies the equation
$E_n\left(R_n(P)\right)=0$, then $R_{n+2}(P)\in
\mathbb{C}[x_1,\dots,\ x_{n+2},\ y_1,\dots,\ y_{n+2}]$ satisfies
the equation $E_{n+2}\left(R_{n+2}(P)\right)=0$.\\
\end{Pro}

\underline{Proof:}\\
We carry out the proof for $B_n$; we proceed likewise for $D_n$.\\
According to remark \ref{grindepnrq}, we may assume that $P$ is
invariant by sign changes.\\
For every $n\geq 2$, we note $Q_n:=R_n(P)$. Then
$$E_{n+1}(Q_{n+1})=R_{n+1}\left(\left(\sum_{k=1}^{n+1}z_ky_k-t_kx_k\right)Q_{n+1}(x_1+z_1,\dots,x_{n+1}+z_{n+1},y_1+t_1,\dots,y_{n+1}+t_{n+1})\right).$$
Let $c_{n+1}:=(1,\ \dots,\ n+1)\in\go{S}_{n+1}$, and $s_{n+1}$ the
$(n+1)-$th sign change, so that $B_{n+1}=\langle s_{n+1},\
c_{n+1}\rangle\cdot B_n$. Then we may write
$$Q_{n+1}=R_{n+1}(P)=\frac{1}{2(n+1)}\sum_{\substack{i=0,1\\ j=1\dots
n+1}}(s_{n+1}^ic_{n+1}^j)\cdot Q_n.$$ Now the polynomial
$(s_{n+1}^ic_{n+1}^j)\cdot Q_n$ contains only the indices
$1,\dots,\ j-1,\ j+1,\dots,\ n+1$,\\therefore
$(z_jy_j-t_jx_j)(s_{n+1}^ic_{n+1}^j)\cdot Q_n$ is in the kernel
of $R_{n+1}$. So, $$E_{n+1}(Q_{n+1})=\frac{1}{2(n+1)}\sum_{\substack{i=0,1\\
j=1\dots n+1}}R_{n+1}\left[\left(\sum_{\substack{k=1\\ k\neq
j}}^{n+1}z_ky_k-t_kx_k\right)\left((s_{n+1}^ic_{n+1}^j)\cdot
Q_n\right)(\mathbf{x}+\mathbf{z},\
\mathbf{y}+\mathbf{t})\right],$$ i.e.
$$\begin{array}{rcl}
  E_{n+1}(Q_{n+1}) & = & \displaystyle{\frac{1}{2(n+1)}\sum_{\substack{i=0,1\\
j=1\dots
n+1}}R_{n+1}\left[(\widetilde{s_{n+1}}^i\widetilde{c_{n+1}}^j)\cdot\left(\left(\sum_{k=1}^{n}z_ky_k-t_kx_k\right)
Q_n(\mathbf{x}+\mathbf{z},\
\mathbf{y}+\mathbf{t})\right)\right]} \\
   & = & \displaystyle{\frac{1}{n+1}\sum_{j=1}^{n+1}R_{n+1}\left[\widetilde{c_{n+1}}^j\cdot\left(\left(\sum_{k=1}^{n}z_ky_k-t_kx_k\right)
Q_n(\mathbf{x}+\mathbf{z},\
\mathbf{y}+\mathbf{t})\right)\right],} \\
\end{array}$$
where the $\widetilde{s_{n+1}}$ and $\widetilde{c_{n+1}}$ act on
the $\mathbf{x},\ \mathbf{y},\ \mathbf{z},\ \mathbf{t}$ as the
$s_{n+1}$ and $c_{n+1}$ act on the $\mathbf{x},\
\mathbf{y}$. (In the $2-$nd equality, we have used the invariance
by sign changes of $P$).\\
By indexing the variables by $\llbr 1,\ n \rrbr$ instead of $\llbr
1,\ n+1 \rrbr\backslash\{j\}$, we see that each of the terms of
this sum is by hypothesis in the kernel of $R_n$, thus in the
kernel of $R_{n+1}$. So,
$E_{n+1}(Q_{n+1})=0$.$\blacksquare$\\

\begin{Cor}$\\$
The sequence $(\dim\, HP_0(B_n))_{n\geq 2}$ is increasing.\\
The sequences $(\dim\, HP_0(D_{2n}))_{n\geq 1}$ and $(\dim\, HP_0(D_{2n+1}))_{n\geq 1}$ are increasing.\\
\end{Cor}

\noindent Proposition \ref{profondconstr} is fundamental for the
construction of the solutions of the equation of
Berest-Etingof-Ginzburg for~$B_n$.\\

\begin{Pro}\label{profondconstr}$\\$
If $R_i(P)$ and $R_j(Q)$ satisfy equation (\ref{equaetingofter}),
and if their sets of indeterminates are \emph{disjoints}, then
$R_{i+j}(PQ)$ satisfies equation (\ref{equaetingofter}).\\
In terms of graphs, it means that if two \emph{disjoints} graphs
satisfy equation (\ref{equaetingofter}), then their union
also satisfies this equation.\\
So it is sufficient that the connected components of a graph
satisfy equation (\ref{equaetingofter}) in order that the graph
itself satisfies it.
\end{Pro}

\underline{Proof:}\\
$\bullet$ Let $R_i(P)$ and $R_j(Q)$ of which the sets of
indeterminates
are disjoints.\\
We denote by \mbox{$E_P:=\{x_k,\ k\in I_P\}\cup\{y_k,\ k\in
I_P\}$} (resp. $E_Q:=\{x_k,\ k\in I_Q\}\cup\{y_k,\ k\in I_Q\}$)
the set of indeterminates of $P$ (resp. $Q$), and we set $n:=i+j$
so that we have $|E_P|=2i,\ |E_Q|=2j$, $I_P\sqcup I_Q=\llbr1,\
n\rrbr$ and $\mathbb{C}[E_P,\ E_Q]=\mathbb{C}[\mathbf{x},\
\mathbf{y}]$. The group $B_i$ (resp. $B_j$) acts only on the
indeterminates
 of $E_P$ (resp. $E_Q$).\\
Let us calculate $R_n(PQ)$:\\
$$\begin{array}{rcl}
  |B_n|R_n(PQ) & = & \displaystyle{\frac{1}{|B_i||B_j|}\sum_{g\in B_i}\sum_{h\in B_j}\sum_{\sigma\in B_n}(gh\sigma)\cdot(PQ)} \\
    & = & \displaystyle{\frac{1}{|B_i||B_j|}\sum_{\sigma\in B_n}\sigma\cdot\left[\sum_{g\in B_i}\sum_{h\in B_j}\underbrace{(h\cdot(g\cdot P))}_{g\cdot P}\underbrace{(h\cdot(g\cdot Q))}_{h\cdot Q}\right]} \\
    & = & \displaystyle{\frac{1}{|B_i||B_j|}\sum_{\sigma\in B_n}\sigma\cdot\left[\left(\sum_{g\in B_i}g\cdot P\right)\left(\sum_{h\in B_j}h\cdot Q\right)\right]} \\
    & = & \displaystyle{\sum_{\sigma\in B_n}\sigma\cdot\left[R_i(P)R_j(Q)\right]} \\
    & = & \displaystyle{|B_n|R_n\left(R_i(P)R_j(Q)\right).} \\
\end{array}$$
\\Hence $R_n(PQ)=R_n\left(R_i(P)R_j(Q)\right)$.\\
$\bullet$ If $R_i(P)$ and $R_j(Q)$ satisfy moreover equation
(\ref{equaetingofter}), we have
$$\begin{array}{rl}
 & \displaystyle{(\mathbf{z}\cdot \mathbf{y}-\mathbf{t}\cdot
\mathbf{x})\left[R_n(PQ)\right](\mathbf{x}+\mathbf{z},\
\mathbf{y}+\mathbf{t})} \\
= & \displaystyle{\frac{1}{|B_n|}\sum_{\sigma\in
B_n}(\mathbf{z}\cdot \mathbf{y}-\mathbf{t}\cdot
\mathbf{x})\left[\sigma\cdot(R_i(P)R_j(Q))\right](\mathbf{x}+\mathbf{z},\
\mathbf{y}+\mathbf{t})} \\
 = & \displaystyle{\frac{1}{|B_n|}\sum_{\sigma\in B_n}\Bigg[\underbrace{\left(\sum_{k\in\sigma^{-1}(I_P)} z_ky_k-t_kx_k\right)[\sigma\cdot R_i(P)](\mathbf{x}+\mathbf{z},\
\mathbf{y}+\mathbf{t})}_{A_{P,\sigma}}\ \underbrace{[\sigma\cdot
R_j(Q)](\mathbf{x}+\mathbf{z},\ \mathbf{y}+\mathbf{t})}_{A_{Q,\sigma}}}\\
 &  \displaystyle{+\
\underbrace{\left(\sum_{k\in\sigma^{-1}(I_Q)}
z_ky_k-t_kx_k\right)[\sigma\cdot R_j(Q)](\mathbf{x}+\mathbf{z},\
\mathbf{y}+\mathbf{t})}_{B_{Q,\sigma}}\ \underbrace{[\sigma\cdot
R_i(P)](\mathbf{x}+\mathbf{z},\
\mathbf{y}+\mathbf{t})}_{B_{P,\sigma}}\Bigg],}
\end{array}$$
where the elements $A_{P,\sigma}$ and $B_{P,\sigma}$ (resp.
$A_{Q,\sigma}$ and $B_{Q,\sigma}$) contain only the
indices of $\sigma^{-1}(I_P)$ (resp. $\sigma^{-1}(I_Q)$).\\
$\bullet$ Let us show that $A_{P,\sigma}A_{Q,\sigma}$ is in the
kernel of $R_n$:
$$\begin{array}{rcl}
  \displaystyle{|B_n|\,R_n(A_{P,\sigma} A_{Q,\sigma})} & = & \displaystyle{\sum_{g\in B_n}g\cdot (A_{P,\sigma} A_{Q,\sigma})} \\
    & = & \displaystyle{\frac{1}{|B_i|}\sum_{h\in B_i}\sum_{g\in hB_n}g\cdot (A_{P,\sigma}A_{Q,\sigma})} \\
        & = & \displaystyle{\frac{1}{|B_i|}\sum_{g\in B_n}g\cdot\left(\sum_{h\in B_i}h\cdot (A_{P,\sigma}A_{Q,\sigma})\right)} \\
    & = & \displaystyle{\frac{1}{|B_i|}\sum_{g\in B_n}g\cdot\left(\left(\sum_{h\in B_i}h\cdot A_{P,\sigma}\right) A_{Q,\sigma}\right)}\\
    & = & \displaystyle{\sum_{g\in B_n}g\cdot\left(\underbrace{R_i(A_{P,\sigma})}_{=0} A_{Q,\sigma}\right)=0.}
\end{array}$$
The last equality is due to the fact that $R_i(P)$ satisfies
equation (\ref{equaetingofter}).\\
Similarly, $B_{P,\sigma}B_{Q,\sigma}$ is in the kernel of $R_n$.
Hence $(\mathbf{z}\cdot \mathbf{y}-\mathbf{t}\cdot
\mathbf{x})\left[R_n(PQ)\right](\mathbf{x}+\mathbf{z},\
\mathbf{y}+\mathbf{t})=0$.$\blacksquare$\\

\noindent So Proposition \ref{profondconstr} gives a way to
construct solutions of equation (\ref{equaetingofter}) for $B_n$
from already known solutions of this equation for $B_m$ with
$m<n$, by taking the disjoint unions of the solution graphs. In
fact, according to Proposition \ref{grindepn2}, if a graph is a
solution for $B_m$, then it is also a solution for
$B_n$ with $n>m$.\\
Thus we may formulate the following conjecture:\\

\begin{Conj}$\\$
For every integer $n\geq 2$, there exists a unique polynomial of
degree $4(n-1)$ in $S^W_{\go{sl}_2}$ (i.e. a linear combination of
graphs with $n-1$ edges) which is solution of equation
(\ref{equaetingofter}). This polynomial is represented by a graph
which is made up with the linear graph without loops and without
isolated vertices, with $n-1$ edges and $n$ vertices, and with
other graphs which may be seen as corrective terms,
and which all have $n-1$ edges and $n$ (non-isolated) vertices.\\
This graph is called the $n-$th simple graph.\\
\end{Conj}

\noindent By some \textsf{Maple} calculations, we determine the list of the first simple graphs:\\

\begin{scriptsize}
\noindent \begin{tabular}{|r|l|}
  \hline
  $B_n$ & $n-$th simple graph \\
  \hline \hline
  $B_2$ & $\xymatrix@R=5pt@C=5pt{
   \bullet\ar@1{-}[r]& \bullet}$ \\ \hline
  $B_3$ & $\xymatrix@R=5pt@C=5pt{
   \bullet\ar@1{-}[r]& \bullet \ar@1{-}[r]& \bullet}$ \\ \hline
$B_4$ & $\xymatrix@R=5pt@C=5pt{
   \bullet\ar@2{-}[r]& \bullet  & \bullet\ar@{-}[r] & \bullet }\ \  -2 \xymatrix@R=5pt@C=5pt{
   \bullet\ar@1{-}[r]& \bullet \ar@1{-}[r] \ar@1{-}[d] & \bullet \\
    & \bullet & }\ \  -10 \xymatrix@R=5pt@C=5pt{
   \bullet\ar@1{-}[r]& \bullet \ar@1{-}[r] & \bullet \ar@1{-}[r] &
   \bullet}$ \\ \hline
$B_5$ & $\xymatrix@R=5pt@C=5pt{ & \bullet & \\
   \bullet\ar@1{-}[r]& \bullet  \ar@{-}[r] \ar@{-}[u] \ar@{-}[d] & \bullet \\ & \bullet & }\ \ \xymatrix@R=5pt@C=5pt{
\\
   -4\ \bullet\ar@2{-}[r]& \bullet \ar@1{-}[r]& \bullet & \bullet\ar@{-}[r] & \bullet }\ \
   \xymatrix@R=5pt@C=5pt{\\
   +28\ \bullet\ar@1{-}[r]& \bullet \ar@1{-}[r] & \bullet \ar@1{-}[r] & \bullet \ar@1{-}[r] & \bullet}\ \
   \xymatrix@R=5pt@C=5pt{\\
  +4\ \bullet\ar@1{-}[r] \ar@1{-}[rd] & \bullet\ar@{-}[d] & \bullet \ar@{-}[r] & \bullet\\ &
  \bullet  &  }\ \   \xymatrix@R=5pt@C=5pt{ \\
   +28\ \bullet\ar@1{-}[r]& \bullet \ar@1{-}[r] & \bullet \ar@1{-}[d]\ar@1{-}[r] & \bullet \\ & & \bullet &
   }$ \\ \hline
$B_6$ &
\begin{tabular}{l}
  $\xymatrix@R=5pt@C=5pt{
   \bullet\ar@3{-}[r]& \bullet  & \bullet\ar@{-}[r] & \bullet & \bullet\ar@{-}[r] & \bullet}\ \
\xymatrix@R=5pt@C=5pt{
   +420\ \bullet\ar@1{-}[r]& \bullet \ar@1{-}[r] & \bullet \ar@1{-}[r] & \bullet\ar@1{-}[r] & \bullet\ar@1{-}[r] & \bullet}\ \
   -70\ \xymatrix@R=5pt@C=5pt{\bullet\ar@1{-}[r]& \bullet \ar@1{-}[r] & \bullet \ar@2{-}[r] & \bullet & \bullet\ar@1{-}[r] &
   \bullet}$ \\

$\xymatrix@R=5pt@C=5pt{ \\ -54\ \bullet\ar@1{-}[r]& \bullet
\ar@2{-}[r] & \bullet  & \bullet\ar@1{-}[r] & \bullet \ar@1{-}[r]
&
     \bullet} \xymatrix@R=5pt@C=5pt{ \\
   +378\ \bullet\ar@1{-}[r]& \bullet \ar@1{-}[r]& \bullet \ar@1{-}[r] & \bullet \ar@1{-}[d]\ar@1{-}[r] & \bullet \\  & & & \bullet & }\ \
   \xymatrix@R=5pt@C=5pt{ \\
   +504\ \bullet\ar@1{-}[r]& \bullet \ar@1{-}[r]& \bullet \ar@1{-}[d]\ar@1{-}[r] & \bullet \ar@1{-}[r] & \bullet \\  & & \bullet & & }\ \
   \xymatrix@R=5pt@C=5pt{ \\
   +\frac{189}{2}\ \bullet\ar@1{-}[r]& \bullet\ar@1{-}[d] \ar@1{-}[r]& \bullet \ar@1{-}[d]\ar@1{-}[r] & \bullet  \\  & \bullet & \bullet & }\ \
$ \\
$\xymatrix@R=5pt@C=5pt{  &\bullet & & \\
   +90\ \bullet\ar@1{-}[r]& \bullet\ar@1{-}[d]\ar@1{-}[u] \ar@1{-}[r]& \bullet \ar@1{-}[r] & \bullet  \\  & \bullet & & }\ \
 \xymatrix@R=5pt@C=5pt{  &\bullet & \bullet & \\
   +\frac{3}{2}\ \bullet\ar@1{-}[r]& \bullet\ar@1{-}[d]\ar@1{-}[u] \ar@1{-}[ru] \ar@1{-}[rd] \\  & \bullet & \bullet& }\ \ \xymatrix@R=5pt@C=5pt{
   \\ +14 &  \bullet\ar@1{-}[r]\ar@1{-}[d]&
\bullet\ar@1{-}[d] & \bullet \ar@1{-}[r] & \bullet  \\
 & \bullet\ar@1{-}[r] & \bullet & }\ \ \xymatrix@R=5pt@C=5pt{\\ +110 & \bullet\ar@1{-}[r]\ar@1{-}[rd]&
\bullet\ar@1{-}[d] & \bullet \ar@1{-}[r] & \bullet \ar@1{-}[r] & \bullet \\
  & & \bullet & }$
\\
$\xymatrix@R=5pt@C=5pt{\\ +56 & \bullet\ar@1{-}[r]\ar@1{-}[rd]&
\bullet\ar@1{-}[d]\ar@1{-}[r] &\bullet & \bullet \ar@1{-}[r] & \bullet  \\
 & & \bullet & }\ \
\xymatrix@R=5pt@C=5pt{\\ -21 & \bullet\ar@1{-}[r]& \bullet
\ar@2{-}[r]
& \bullet\ar@1{-}[r]& \bullet  & \bullet\ar@1{-}[r] & \bullet }\ \  \xymatrix@R=5pt@C=5pt{ \\
   -15\ \bullet\ar@1{-}[r]& \bullet\ar@2{-}[d] \ar@1{-}[r]& \bullet  & \bullet \ar@1{-}[r] & \bullet \\  & \bullet & &  &
   }$ \\
\end{tabular}
 \\ \hline
\end{tabular}\\
$\\$ \end{scriptsize}

\noindent The number of graphs that we may construct from the
simple graphs by using Proposition \ref{profondconstr} equals the
number of linear graphs without loops and without isolated
vertices of Proposition \ref{grapheslin\'eaires}, since the graphs
constructed like this are obtained by substituting, to certain
linear graphs without loops and without isolated vertices, a
linear combination of graphs containing this graph and other
graphs (possibly non-linear or with loops), but
having the same number of vertices and the same number of edges.\\

\begin{Conj}$\\$
For every integer $n\geq 2$, the $\pi(n)$ graphs constructed
according to the process defined above are the only solutions of
 equation (\ref{equaetingofter}).\\
\end{Conj}

\noindent As for $D_n$, we may write, with the help of
\textsf{Maple} and of Proposition \ref{lienBnDnbis}, the solutions
of degree lower than $4(n-1)$ of equation (\ref{equaetingofter})
for $n\in\{2,\,3,\,4\}$. We conjecture that this equation has no
solution of degree strictly
higher than $4(n-1)$.\\

\begin{scriptsize}
\noindent \begin{tabular}{|r|l|}
  \hline
  $D_n$ & Solutions of equation (\ref{equaetingofter})\\
  \hline \hline
  $D_2$ & \begin{tabular}{l}
    $1$ \\
  \end{tabular} \\ \hline
  $D_3$ & \begin{tabular}{l}
    $1$ \\
  \end{tabular} \\ \hline
$D_4$ & \begin{tabular}{l|l|l} $1$ & $\xymatrix@R=5pt@C=5pt{
   \bullet\ar@1{-}[r]& \bullet}$ & $\xymatrix@R=5pt@C=5pt{
   \bullet\ar@1{-}[r]& \bullet  & \bullet\ar@{-}[r] & \bullet }\ \  -2 \xymatrix@R=5pt@C=5pt{
   \bullet\ar@1{-}[r]& \bullet \ar@1{-}[r] & \bullet}$\\
\end{tabular} \\ \hline
\end{tabular}\\
\end{scriptsize}$\\$

\section{\textsf{Study of $B_2-C_2$, $D_2=A_1\times A_1$, $B_3-C_3$, and $D_3=A_3$}}

\noindent Let us now use the results of the general case
previously established by applying them to the cases of $B_2$,
$D_2$, $B_3$ and $D_3$. It will enable us to calculate explicitly
the dimension of their $0-$th space of Poisson homology, and so to
treat entirely
the rank $3$, and at the same time to verify the conjecture of J. Alev for this rank.\\

\subsection{\textsf{Study of $B_2$ and $D_2$}}

\noindent The computation of the dimension of the $0-$th space of
Poisson homology in the case of $B_2$ was made by J. Alev and L.
Foissy in \cite{AF06}. Here, we rediscover this result by another
method. We will then deduce the dimension of the $0-$th space of
Poisson homology
in the case of $D_2$.\\

\begin{Pro}\label{proB2}$\\$
For $B_2$, the space of solutions of equation
(\ref{equaetingofter}) is the plane generated by the polynomials
$1$ and \mbox{$(x_1y_2-y_1x_2)^2$.} So, the dimension of the
$0-$th space of Poisson homology of $B_2$ is
\fbox{$\dim(HP_0(B_2))=2$}.\\
The elements $1$ and $(x_1y_2-y_1x_2)^2$ do not belong to $\{S^W,\
S^W\}$.
\end{Pro}

\underline{Proof:}\\
$\bullet$ Let $X=x_1y_2-y_1x_2$. We have $\mathbb{C}[X]\subset
S_{\go{sl}_2}$. Now, according to Proposition \ref{alev}, we have
$\dim\left(S(2l)_{\go{sl}_2}\right)=1$ therefore the Poincar\'e
series of $S_{\go{sl}_2}$ is $\frac{1}{1-z^2}$. This series is
exactly the Poincar\'e series of $\mathbb{C}[X]$, so
$S_{\go{sl}_2}=\mathbb{C}[X]$. Consequently,
$S_{\go{sl}_2}^W=\mathbb{C}[X]^W=R_2(\mathbb{C}[X])=\mathbb{C}[X^2].$\\
$\bullet$ According to Proposition \ref{deg0&4}, the elements $1$
and $X^2$ do not belong to $\{S^W,\ S^W\}$. We show by a direct
calculation (for example with \textsf{Maple}; see section 4)
 that the element $X^2$ satisfies equation (\ref{equaetingofter}).\\
$\bullet$ In order to have the asserted result, it is sufficient,
according to Corollary \ref{profond}, to show that for every
\mbox{$j\in \mathbb{N}\backslash\{0,\ 1\}$}, the element $X^{2j}$
does not satisfy equation (\ref{equaetingofter}), i.e.
$E_2(X^{2j}) \neq 0$. To do this, it is sufficient, according to
remark \ref{rqev}, to show that $E'_2(X^{2j}) \neq 0$. From the
expression
$$E_{int}^2(X^{2j})=z_1R_2\big(y_1((x_1+z_1)y_2-(y_1+t_1)(x_2+z_2))^{2j}\big)+z_2R_2\big(y_2((x_1+z_1)y_2-(y_1+t_1)(x_2+z_2))^{2j}\big),$$
we deduce
\begin{center}
$\displaystyle{E'_2(X^{2j})=\frac{1}{8}\,z_1\sum_{sign\ changes}
y_1(-(y_1+t_1)z_2)^{2j}+\frac{1}{8}\,z_2\sum_{sign\ changes}
y_1(z_1y_1-t_1z_2)^{2j}},$
\end{center}
i.e.
\begin{center}
$E'_2(X^{2j})=\frac{1}{4}\big[
z_1y_1z_2^{2j}(y_1+t_1)^{2j}-z_1y_1z_2^{2j}(-y_1+t_1)^{2j}+z_2y_1(z_1y_1-t_1z_2)^{2j}-z_2y_1(-z_1y_1-t_1z_2)^{2j}
\big].$
\end{center}
\vspace{.2cm}So the term of the type  $\alpha
z_1t_1y_1^{2j}z_2^{2j}$
 is
$\frac{1}{4}z_1y_1z_2^{2j}(C_{2j}^1y_1^{2j-1}t_1)-\frac{1}{4}z_1y_1z_2^{2j}(-C_{2j}^1y_1^{2j-1}t_1)$,
and we find $\alpha=j\neq 0$. This shows that $X^{2j}$ does not
satisfy equation~(\ref{equaetingofter}).$\blacksquare$

\begin{Pro}\label{proD2}$\\$
For $D_2=A_1\times A_1$, the space of the solutions of equation
(\ref{equaetingofter}) is the one-dimensional space generated by
$1$. So the dimension of the $0-$th space of Poisson homology of
$D_2$ is $1$, i.e.
\fbox{$\dim(HP_0(D_2))=1$}.\\
The element $1$ is the only element which does not belong to
$\{S^W,\ S^W\}$.
\end{Pro}

\underline{Proof:}\\
The method is the same as the one for $B_2$: letting
$X=x_1y_2-y_1x_2$, we have $S_{\go{sl}_2}^W=\mathbb{C}[X^2]$.
According to Proposition \ref{deg0&4}, the element $1$ does not
belong to $\{S^W,\ S^W\}$. We show by a calculation (for example
with \textsf{Maple}; see section 4) that the element $X^2$ does
not satisfy
equation (\ref{equaetingofter}).\\
So it is sufficient, according to Corollary \ref{profond}, to show
that for every \mbox{$j\in \mathbb{N}\backslash\{0,\ 1\}$},
$E'_2(X^{2j}) \neq 0$. Now this expression is the same as the one
obtained for $B_2$; therefore it is not zero.$\blacksquare$

\subsection{\textsf{Study of $B_3$ - Vectors of highest weight $0$}}

\noindent In order to study $B_3$, we begin making by making
the vectors of highest weight $0$ explicit.\\

\begin{Pro}$\\$
We set $X=x_1y_2-y_1x_2,\ Y=x_2y_3-y_2x_3$ and $Z=x_3y_1-y_3x_1$.
\\Then $S^{B_3}_{\go{sl}_2}$ is the image of $\mathbb{C}[X,\
Y,\ Z]$ by $R_3$.
\end{Pro}

\underline{Proof:} according to Proposition \ref{pfaff},
$\mathbb{C}[X,\ Y,\ Z]= S_{\go{sl}_2}$, thus $\mathbb{C}[X,\ Y,\
Z]^{B_3}=S_{\go{sl}_2}^{B_3}$.\\

\begin{Rq}\label{vphpB3} $\\$
The only monomials $X^iY^jZ^k\ (i,\ j,\ k)\in \mathbb{N}^3$ of
which the image by $R_3$ is not zero are \\
$\triangleright$ the monomials of the form $X^iY^jZ^k$ with $i,\
j,\ k$ even,\\
$\triangleright$ the monomials of the form $X^iY^jZ^k$ with $i,\
j,\ k$ odd and all distinct. For these monomials, it is even sufficient to take $i<j<k$.\\
So the vector of highest weigh $0$ of smallest degree not multiple
of $4$ has degree $2(1+3+5)=18$.\\
\end{Rq}

\begin{Pro}\label{deg0&4&8}$\\$
The elements $1$ and $R_3(X^2)$ do not belong to
$\{S^W,\ S^W\}$.\\
The elements $1$, $R_3(X^2)$ and $R_3(X^2Y^2)$ satisfy equation
(\ref{equaetingofter}).\\
So, $\dim(HP_0(B_3))\geq 3$.
\end{Pro}

\underline{Proof:}\\
We know according to Proposition \ref{deg0&4} that the elements
$1$
and $R_3(X^2)$ do not belong to $\{S^W,\ S^W\}$.\\
To show that the elements $R_3(X^2)$ and $R_3(X^2Y^2)$ satisfy
equation (\ref{equaetingofter}), we make a calculation with
\textsf{Maple}:
 see section 4.$\blacksquare$\\

\noindent In order to prove the equality $\dim(HP_0(B_3))=3$, we
will show that the elements of $S^{B_3}_{\go{sl}_2}$ which do not
belong to the space $\langle 1,\ R_3(X^2),\ R_3(X^2Y^2) \rangle$
do not satisfy equation (\ref{equaetingofter}). It is sufficient,
according to remark \ref{rqev}, to show that these elements do not
satisfy equation (\ref{equaetingofsubs}). It is the
aim of the following section.\\

\subsection{\textsf{Study of $B_3$ - Equation of Berest-Etingof-Ginzburg}}

\noindent To each polynomial of the form $P:=R_3(X^iY^jZ^k)$
specified in remark \ref{vphpB3}, we associate a monomial $M_P$
which appears with a non-zero coefficient in equation
(\ref{equaetingofsubs}), and such that if $P_1$ and $P_2$ are two
distinct polynomials, then the monomial $M_{P_1}$ does not appear
in $E'_n(P_2)$ and the monomial $M_{P_2}$ does not appear in
$E'_n(P_1)$, where $E'_n(P)$ denotes equation
(\ref{equaetingofsubs}).\\

To the polynomial  $R_3(X^{2j})$ (with $j\geq 2$), we associate the monomial $M_j=z_1t_1y_1^{2j}z_3^{2j}$.\\

To the polynomial  $R_3(X^{2j}Y^{2l})$ (with $1\leq l\leq j$ and
$j\geq
2$), we associate $M_{j,l}=z_1t_1y_1^{2j+2l}z_3^{2j}z_2^{2l}.$\\

To the polynomial  $R_3(X^{2j}Y^{2l}Z^{2k})$ (with $1\leq l\leq
k\leq j$),
we associate $M_{j,k,l}=z_1t_1y_1^{2j+2l}z_3^{2j+2k}z_2^2t_1^{2k}z_1^{2l-2}.$\\

To the polynomial  $R_3(X^{2i+1}Y^{2j+1}Z^{2k+1})$ (with $0\leq i<
j< k$), we associate
$\widetilde{M}_{k,j,i}=z_1t_1y_1^{2i+2k+2}z_3^{2j+2k+2}z_2^2t_1^{2j+1}z_1^{2i-1}.$\\

\subsubsection{\textsf{First step}}

\noindent For every polynomial $P$, we calculate the
coefficient of the monomial $M_P$ which appears in equation $E'_3(P)$.\\

\textbf{Case 1.} $P=R_3(X^{2j})=\frac{1}{3}(X^{2j}+Y^{2j}+Z^{2j})$
(with $j\geq 2$):\\
We have
\begin{center}
$|W|E_{int}^3(P)=\frac{|W|}{3}R_3\Big[ (z_1y_1+z_2y_2+z_3y_3)\big[
((z_1y_2-(y_1+t_1)z_2)^{2j}+(z_2y_3-y_2z_3)^{2j}+(z_3(y_1+t_1)-y_3z_1)^{2j}\big]
\Big],$
\end{center}
Thus
\begin{center}
\begin{small}
$\begin{array}{c}
\displaystyle{|W|E_{int}^3(P)=y_2\,(\,\dots)+y_3\,(\,\dots)+\frac{2}{3}\sum_{sign\
changes}z_1y_1\big[ ((y_1+t_1)z_2)^{2j}+(z_3(y_1+t_1))^{2j}\big]}
\\
\displaystyle{+\frac{2}{3}\sum_{sign\ changes}z_2y_1\big[
(z_1y_1-t_1z_2)^{2j}+(y_1z_3)^{2j}+(t_1z_3)^{2j}\big]
+\frac{2}{3}\sum_{sign\ changes}z_3y_1\big[
(t_1z_2)^{2j}+(z_2y_1)^{2j}+(z_3t_1-y_1z_1)^{2j}\big]}.
\end{array}$
\end{small}
\end{center}
Therefore,
\begin{center}
\begin{small}
\begin{equation}\label{equagraphe1}
\begin{array}{rcl} |W|E'_3(P) & = &
\frac{8}{3}\Big[z_1y_1\big[
((y_1+t_1)z_2)^{2j}+(z_3(y_1+t_1))^{2j}\big]-z_1y_1\big[
((-y_1+t_1)z_2)^{2j}+(z_3(-y_1+t_1))^{2j}\big]\Big]
\\
 & & +\frac{8}{3}\Big[z_2y_1\big[
(z_1y_1-t_1z_2)^{2j}+\underline{(y_1z_3)^{2j}}+\underline{(t_1z_3)^{2j}}\big]-z_2y_1\big[
(-z_1y_1-t_1z_2)^{2j}+\underline{(y_1z_3)^{2j}}+\underline{(t_1z_3)^{2j}}\big]\Big]\\
 & & +\frac{8}{3}\Big[z_3y_1\big[
\underline{(t_1z_2)^{2j}}+\underline{(z_2y_1)^{2j}}+(z_3t_1-y_1z_1)^{2j}\big]-z_3y_1\big[
\underline{(t_1z_2)^{2j}}+\underline{(z_2y_1)^{2j}}+(z_3t_1+y_1z_1)^{2j}\big]\Big],
\end{array}
\end{equation}
\end{small}
\end{center}
where the underlined terms cancel out.\\
The monomial $M_j=z_1t_1y_1^{2j}z_3^{2j}$ appears only in the
first line of the last expression, and its coefficient is
$\frac{1}{|W|}\frac{8}{3}\big[2C_{2j}^1+2C_{2j}^1\big]=\frac{32}{3}j$.\\
So, \fbox{the coefficient of $M_j$ in $E'_3(P)$ is
$\frac{2}{9}j$}.\\

\textbf{Case 2.} $P=R_3(X^{2j}Y^{2l})$ (with $1\leq l\leq j$ and $j\geq 2$):\\

\noindent We have
\begin{center}
\begin{small}
$\begin{array}{c} |W|E_{int}^3(P)=\frac{|W|}{6}R_3\Big[
(z_1y_1+z_2y_2+z_3y_3)\\
\big[
((z_1y_2-(y_1+t_1)z_2)^{2j}(z_2y_3-y_2z_3)^{2l}+(z_2y_3-y_2z_3)^{2j}(z_3(y_1+t_1)-y_3z_1)^{2l}+(z_3(y_1+t_1)-y_3z_1)^{2j}(z_1y_2-(y_1+t_1)z_2)^{2l}\\
((z_1y_2-(y_1+t_1)z_2)^{2l}(z_2y_3-y_2z_3)^{2j}+(z_2y_3-y_2z_3)^{2l}(z_3(y_1+t_1)-y_3z_1)^{2j}+(z_3(y_1+t_1)-y_3z_1)^{2l}(z_1y_2-(y_1+t_1)z_2)^{2j}\big]
\Big],
\end{array}$
\end{small}
\end{center}
\begin{center}
\begin{small}
\begin{equation}\label{equagraphe2}
\begin{array}{rlc}\textrm{Thus}\ \ |W|E'_3(P) & = &  \frac{8}{6}\Big[
z_1y_1\big[(z_3(y_1+t_1))^{2j}((y_1+t_1)z_2)^{2l}
\big]-z_1y_1\big[(z_3(-y_1+t_1))^{2j}((-y_1+t_1)z_2)^{2l} \big]
\\
 & & +z_2y_1\big[ (z_1y_1-t_1z_2)^{2j}(y_1z_3)^{2l}+\underline{(y_1z_3)^{2j}(z_3t_1)^{2l}}+(z_3t_1)^{2j}(z_1y_1-t_1z_2)^{2l}
 \big]-\dots\\
 & & +z_3y_1\big[
\underline{(t_1z_2)^{2j}(z_2y_1)^{2l}}+(z_2y_1)^{2j}(z_3t_1-y_1z_1)^{2l}+(z_3t_1-y_1z_1)^{2j}(t_1z_2)^{2l}\big]-\dots\\
 & & +z_1y_1\big[(z_3(y_1+t_1))^{2l}((y_1+t_1)z_2)^{2j}
\big]-z_1y_1\big[(z_3(-y_1+t_1))^{2l}((-y_1+t_1)z_2)^{2j} \big]
\\
  & & +z_2y_1\big[ (z_1y_1-t_1z_2)^{2l}(y_1z_3)^{2j}+\underline{(y_1z_3)^{2l}(z_3t_1)^{2j}}+(z_3t_1)^{2l}(z_1y_1-t_1z_2)^{2j}
 \big]-\dots\\
 & & +z_3y_1\big[
\underline{(t_1z_2)^{2l}(z_2y_1)^{2j}}+(z_2y_1)^{2l}(z_3t_1-y_1z_1)^{2j}+(z_3t_1-y_1z_1)^{2l}(t_1z_2)^{2j}\big]-\dots
\Big],
\end{array}
\end{equation}
\end{small}
\end{center}
where on each line, the suspension points stand for the image of
the expression by the sign change $y_1\mapsto -y_1$. As in the
preceding case, the underlined terms cancel out.\\
$\triangleright$ If $j\neq l$ and $l\neq 1$, the monomial
$M_{j,l}$ appears only in the first line of the preceding
expression and its coefficient in this
expression is $\frac{32}{6}(j+l)$.\\
$\triangleright$ If $j\neq l$ and $l=1$, the monomial $M_{j,l}$
also appears in the $5-$th line of (\ref{equagraphe2}) and its coefficient is $\frac{32}{6}(j+l-l)$.\\
$\triangleright$ If $j=l$, we have $l\neq 1$ (because $j\geq 2$).
Then $M_{j,l}$ also appears in the fourth
line and its coefficient is thus $\frac{64}{6}(j+l)$.\\

\noindent The following table collects the coefficients of the monomial $M_{j,l}$ in $E'_3(P)$.\\
\begin{center}
\begin{tabular}{|c|c|c|c|}
  \hline
  $(j,\,l)$ & $j\neq l$ and $l\neq 1$ & $j\neq l$ and $l=1$ & $j=l$ \\
  \hline
  Coefficient & $(j+l)/9$ & $j/9$ & $4j/9$ \\
  \hline
\end{tabular}
\end{center}$\\$

\textbf{Case 3.}  $P=R_3(X^{2j}Y^{2l}Z^{2k})$ (with $1\leq l\leq
k\leq j$):\\

\noindent we have
\begin{center}
\begin{small}
$\begin{array}{c}
\displaystyle{|W|E_{int}^3(P)=\frac{|W|}{6}R_3\Big[
(z_1y_1+z_2y_2+z_3y_3)\big[
((z_1y_2-(y_1+t_1)z_2)^{2j}(z_2y_3-y_2z_3)^{2l}(z_3(y_1+t_1)-y_3z_1)^{2k}+\sum_{\substack{permutations\\
of\ (j,l,k)}}\dots\big]\Big]},
\end{array}$
\end{small}
\end{center}
\begin{center}
\begin{small}
$\begin{array}{c} \textrm{So}\ \ \ \ \ \ \ \  |W|E'_3(P)=\frac{8}{6}\displaystyle{\sum_{\substack{permutations\\
of\ (j,l,k)}}\Big[
z_2y_1(z_1y_1-t_1z_2)^{2j}(y_1z_3)^{2l}(z_3t_1)^{2k}-z_2y_1(-z_1y_1-t_1z_2)^{2j}(y_1z_3)^{2l}(z_3t_1)^{2k}}\\
+z_3y_1(t_1z_2)^{2j}(z_2y_1)^{2l}(z_3t_1-y_1z_1)^{2k}-z_3y_1(t_1z_2)^{2j}(z_2y_1)^{2l}(z_3t_1+y_1z_1)^{2k}\Big],
\end{array}$
\end{small}
\end{center}
We denote by $\alpha_{j,k,l}$ the coefficient of the monomial
$M_{j,k,l}$ in the preceding expression, and we distinguish 4 cases:\\

\noindent $\triangleright$ If $l=k=j$, we have
$M_{j,j,j}=z_1t_1z_1^{2j-2}y_1^{4j}z_3^{4j}z_2^2t_1^{2j}$ and
$\alpha_{j,j,j}=-32j$.\\
\noindent $\triangleright$ If $l=k<j$, we have
$M_{j,l,l}=z_1t_1z_1^{2l-2}y_1^{2j+2l}z_3^{2j+2l}z_2^2t_1^{2l}$,
and
\begin{center}
\begin{small}
$\begin{array}{c} |W|E'_3(P)=\frac{8}{6}\Big[
z_2y_1(z_1y_1-t_1z_2)^{2j}(y_1z_3)^{2l}(z_3t_1)^{2l}-\dots
+z_3y_1(t_1z_2)^{2j}(z_2y_1)^{2l}(z_3t_1-y_1z_1)^{2l}-\dots\\
+z_2y_1(z_1y_1-t_1z_2)^{2l}(y_1z_3)^{2j}(z_3t_1)^{2l}-\dots
+z_3y_1(t_1z_2)^{2l}(z_2y_1)^{2j}(z_3t_1-y_1z_1)^{2l}-\dots\\
+z_2y_1(z_1y_1-t_1z_2)^{2l}(y_1z_3)^{2l}(z_3t_1)^{2j}-\dots
+z_3y_1(t_1z_2)^{2l}(z_2y_1)^{2l}(z_3t_1-y_1z_1)^{2j}-\dots\Big],
\end{array}$
\end{small}
\end{center}

\noindent where on each line, the suspension points stand for the
image of the expression by the sign change $y_1\mapsto
-y_1$. So the searched coefficient is $\alpha_{j,l,l}=-\frac{32}{3}l$.\\

\noindent $\triangleright$ If $l<k=j$, we have
$M_{j,j,l}=z_1t_1z_1^{2l-2}y_1^{2j+2l}z_3^{4j}z_2^2t_1^{2j}$. Similarly, we find \mbox{$\alpha_{j,j,l}=-32j$.}\\

\noindent $\triangleright$ If $l<k<j$, we have
\begin{center}
\begin{small}
\begin{equation}\label{equagraphe3}
\begin{array}{rcl} |W|E'_3(P) & = & \frac{8}{6}\Big[
z_2y_1(z_1y_1-t_1z_2)^{2j}(y_1z_3)^{2l}(z_3t_1)^{2k}-\dots\\
 & & +z_3y_1(t_1z_2)^{2j}(z_2y_1)^{2l}(z_3t_1-y_1z_1)^{2k}-\dots\\
 & & +z_2y_1(z_1y_1-t_1z_2)^{2j}(y_1z_3)^{2k}(z_3t_1)^{2l}-\dots\\
 & & +z_3y_1(t_1z_2)^{2j}(z_2y_1)^{2k}(z_3t_1-y_1z_1)^{2l}-\dots\\
 & & +z_2y_1(z_1y_1-t_1z_2)^{2k}(y_1z_3)^{2j}(z_3t_1)^{2l}-\dots\\
 & & +z_3y_1(t_1z_2)^{2k}(z_2y_1)^{2j}(z_3t_1-y_1z_1)^{2l}-\dots\\
 & & +z_2y_1(z_1y_1-t_1z_2)^{2l}(y_1z_3)^{2j}(z_3t_1)^{2k}-\dots\\
 & & +z_3y_1(t_1z_2)^{2l}(z_2y_1)^{2j}(z_3t_1-y_1z_1)^{2k}-\dots\\
 & & +z_2y_1(z_1y_1-t_1z_2)^{2k}(y_1z_3)^{2l}(z_3t_1)^{2j}-\dots\\
 & & +z_3y_1(t_1z_2)^{2k}(z_2y_1)^{2l}(z_3t_1-y_1z_1)^{2j}-\dots\\
 & & +z_2y_1(z_1y_1-t_1z_2)^{2l}(y_1z_3)^{2k}(z_3t_1)^{2j}-\dots\\
 & & +z_3y_1(t_1z_2)^{2l}(z_2y_1)^{2k}(z_3t_1-y_1z_1)^{2j}-\dots\Big],
\end{array}
\end{equation}
\end{small}
\end{center}
where, again, the suspension points stand for the image of the
expression by the sign change $y_1\mapsto
-y_1$.\\
The searched monomial appears only on the $7-$th line, and its
coefficient is $\alpha_{j,k,l}=-\frac{32}{6}l$.\\

\noindent Finally we collect in the following table the
coefficients of the monomial $M_{j,k,l}$ in $E'_3(P)$.\\

\begin{center}
\begin{tabular}{|c|c|c|c|c|}
  \hline
  $(j,\,k,\,l)$ & $l=k=j$ & $l=k<j$ & $l<k=j$ & $l<k<j$\\
  \hline
  Coefficient & $-2l/3$ & $-2l/9$ & $-2l/9$ & $-l/9$ \\
  \hline
\end{tabular}
\end{center}$\\$

\textbf{Case 4.}  $P=R_3(X^{2i+1}Y^{2j+1}Z^{2k+1})$ (with $0\leq
i< j< k$): we have
\begin{center}
\begin{scriptsize}
$\begin{array}{c}
\displaystyle{|W|E_{int}^3(P)=\frac{|W|}{6}R_3\Big[
(z_1y_1+z_2y_2+z_3y_3)\big[
((z_1y_2-(y_1+t_1)z_2)^{2i+1}(z_2y_3-y_2z_3)^{2j+1}(z_3(y_1+t_1)-y_3z_1)^{2k+1}+\sum_{\substack{permutations\\
of\ (i,j,k)}}\dots\big]\Big]},
\end{array}$
\end{scriptsize}
\end{center}

\begin{small}
\begin{equation}\label{equagraphe4}
\begin{array}{rcl} \textrm{So}\ \ |W|E'_3(P)&=&\frac{8}{6}\displaystyle{\sum_{\substack{permutations\\
of\ (j,j,k)}}\Big[
z_2y_1(z_1y_1-t_1z_2)^{2i+1}(y_1z_3)^{2j+1}(z_3t_1)^{2k+1}+z_2y_1(-z_1y_1-t_1z_2)^{2i+1}(y_1z_3)^{2j+1}(z_3t_1)^{2k+1}}\\
 &
 &+z_3y_1(t_1z_2)^{2i+1}(z_2y_1)^{2j+1}(z_3t_1-y_1z_1)^{2k+1}+z_3y_1(t_1z_2)^{2i+1}(z_2y_1)^{2j+1}(z_3t_1+y_1z_1)^{2k+1} \Big]\\

  &=&z_2y_1(z_1y_1-t_1z_2)^{2i+1}(y_1z_3)^{2j+1}(z_3t_1)^{2k+1}+z_2y_1(-z_1y_1-t_1z_2)^{2i+1}(y_1z_3)^{2j+1}(z_3t_1)^{2k+1}\\
 & &+z_3y_1(t_1z_2)^{2i+1}(z_2y_1)^{2j+1}(z_3t_1-y_1z_1)^{2k+1}+z_3y_1(t_1z_2)^{2i+1}(z_2y_1)^{2j+1}(z_3t_1+y_1z_1)^{2k+1}\\

 & &+z_2y_1(z_1y_1-t_1z_2)^{2i+1}(y_1z_3)^{2k+1}(z_3t_1)^{2j+1}+z_2y_1(-z_1y_1-t_1z_2)^{2i+1}(y_1z_3)^{2k+1}(z_3t_1)^{2j+1}\\
 & &+z_3y_1(t_1z_2)^{2i+1}(z_2y_1)^{2k+1}(z_3t_1-y_1z_1)^{2j+1}+z_3y_1(t_1z_2)^{2i+1}(z_2y_1)^{2k+1}(z_3t_1+y_1z_1)^{2j+1}\\

& &+z_2y_1(z_1y_1-t_1z_2)^{2j+1}(y_1z_3)^{2i+1}(z_3t_1)^{2k+1}+z_2y_1(-z_1y_1-t_1z_2)^{2j+1}(y_1z_3)^{2i+1}(z_3t_1)^{2k+1}\\
& &+z_3y_1(t_1z_2)^{2j+1}(z_2y_1)^{2i+1}(z_3t_1-y_1z_1)^{2k+1}+z_3y_1(t_1z_2)^{2j+1}(z_2y_1)^{2i+1}(z_3t_1+y_1z_1)^{2k+1}\\

 & &+z_2y_1(z_1y_1-t_1z_2)^{2k+1}(y_1z_3)^{2j+1}(z_3t_1)^{2i+1}+z_2y_1(-z_1y_1-t_1z_2)^{2k+1}(y_1z_3)^{2j+1}(z_3t_1)^{2i+1}\\
 & &+z_3y_1(t_1z_2)^{2k+1}(z_2y_1)^{2j+1}(z_3t_1-y_1z_1)^{2i+1}+z_3y_1(t_1z_2)^{2k+1}(z_2y_1)^{2j+1}(z_3t_1+y_1z_1)^{2i+1}\\

 & &+z_2y_1(z_1y_1-t_1z_2)^{2j+1}(y_1z_3)^{2k+1}(z_3t_1)^{2i+1}+z_2y_1(-z_1y_1-t_1z_2)^{2j+1}(y_1z_3)^{2k+1}(z_3t_1)^{2i+1}\\
 & &+z_3y_1(t_1z_2)^{2j+1}(z_2y_1)^{2k+1}(z_3t_1-y_1z_1)^{2i+1}+z_3y_1(t_1z_2)^{2j+1}(z_2y_1)^{2k+1}(z_3t_1+y_1z_1)^{2i+1}\\

 & &+z_2y_1(z_1y_1-t_1z_2)^{2k+1}(y_1z_3)^{2i+1}(z_3t_1)^{2j+1}+z_2y_1(-z_1y_1-t_1z_2)^{2k+1}(y_1z_3)^{2i+1}(z_3t_1)^{2j+1}\\
 & &+z_3y_1(t_1z_2)^{2k+1}(z_2y_1)^{2i+1}(z_3t_1-y_1z_1)^{2j+1}+z_3y_1(t_1z_2)^{2k+1}(z_2y_1)^{2i+1}(z_3t_1+y_1z_1)^{2j+1}\\
\end{array}
\end{equation}
\end{small}

\noindent In order to find the monomials of the form $\alpha
\widetilde{M}_{k,j,i}$, we are first interested in the exponent of
$z_3$, then in the one of $y_1$. So it remains only the lines 1
and 3, then only the line 3. The coefficient of
$\widetilde{M}_{k,j,i}$ in $|W|E'_3(P)$ is
$-\frac{16}{6}(2i+1)$.\\
So, \fbox{the coefficient of $\widetilde{M}_{k,j,i}$ in $E'_3(P)$
is
$-\frac{1}{18}(2i+1)$}.\\

\subsubsection{\textsf{Second step}}

\noindent For two distinct polynomials $P_1$ and $P_2$, of the
same degree greater or equal to $12$, we show that $M_{P_1}$
does not appear in $E'(P_2)$.\\

\textbf{Case 1.}  $P=R_3(X^{2j})$ ($j\geq 3$): according to
(\ref{equagraphe1}), no term of
$E'_3(P)$ contains at the same time~$z_1,\ z_2,\ z_3$.\\
Now if $l+k=j$ with $1\leq k\leq l$ and $l\geq 2$, then
$M_{l,k}=z_1t_1y_1^{2l+2k}z_3^{2l}z_2^{2k}$, therefore $M_{l,k}$
does not appear in $E'_3(P)$.\\
Similarly, if $m+l+k=j$ with $0<k\leq l\leq m$, we have
$M_{m,l,k}=z_1t_1y_1^{2m+2k}z_3^{2m+2l}z_2^2t_1^{2l}z_1^{2k-2}$,
therefore $M_{m,l,k}$ does not appear in $E'_3(P)$.\\

\textbf{Case 2.}  $P=R_3(X^{2j}Y^{2l})$ (with $1\leq l\leq j$ and
$j\geq 2$): equation (\ref{equagraphe2}) may be written
\begin{center}
\begin{small}
\begin{equation}\label{equagraphe2bis}
\begin{array}{rcl} \frac{6}{8}|W|E'_3(P) & = &
 z_1y_1(z_3(y_1+t_1))^{2j}((y_1+t_1)z_2)^{2l}-z_1y_1(z_3(-y_1+t_1))^{2j}((-y_1+t_1)z_2)^{2l}\\
 & & +z_1y_1(z_3(y_1+t_1))^{2l}((y_1+t_1)z_2)^{2j}-z_1y_1(z_3(-y_1+t_1))^{2l}((-y_1+t_1)z_2)^{2j}\\
 & & +z_2y_1(z_1y_1-t_1z_2)^{2j}(y_1z_3)^{2l}+z_2y_1(z_3t_1)^{2j}(z_1y_1-t_1z_2)^{2l}\\
 & & -z_2y_1(-z_1y_1-t_1z_2)^{2j}(y_1z_3)^{2l}-z_2y_1(z_3t_1)^{2j}(-z_1y_1-t_1z_2)^{2l}\\
 & & +z_2y_1(z_1y_1-t_1z_2)^{2l}(y_1z_3)^{2j}+z_2y_1(z_3t_1)^{2l}(z_1y_1-t_1z_2)^{2j}\\
 & & -z_2y_1(-z_1y_1-t_1z_2)^{2l}(y_1z_3)^{2j}-z_2y_1(z_3t_1)^{2l}(-z_1y_1-t_1z_2)^{2j}\\
 & & +z_3y_1(z_2y_1)^{2j}(z_3t_1-y_1z_1)^{2l}+z_3y_1(z_3t_1-y_1z_1)^{2j}(t_1z_2)^{2l}\\
 & & -z_3y_1(z_2y_1)^{2j}(z_3t_1+y_1z_1)^{2l}-z_3y_1(z_3t_1+y_1z_1)^{2j}(t_1z_2)^{2l}\\
 & & +z_3y_1(z_2y_1)^{2l}(z_3t_1-y_1z_1)^{2j}+z_3y_1(z_3t_1-y_1z_1)^{2l}(t_1z_2)^{2j}\\
 & &
 -z_3y_1(z_2y_1)^{2l}(z_3t_1+y_1z_1)^{2j}-z_3y_1(z_3t_1+y_1z_1)^{2l}(t_1z_2)^{2j}.
\end{array}
\end{equation}
\end{small}
\end{center}
$\bullet$ If $m=j+l$, then $M_m=z_1t_1y_1^{2m}z_3^{2m}$ does not
appear in $E(P)$. Indeed, the largest exponent
of $z_3$ in $E(P)$ is $\leq 2j+1<2m$.\\

\noindent $\bullet$ If $m+p=j+l$, with $1\leq p\leq m$, $m\geq 2$,
$m\neq j $ and $p \neq l$, then
$M_{m,p}=z_1t_1y_1^{2m+2p}z_3^{2m}z_2^{2p}$ does not appear in
$E(P)$. In fact,\\
$\triangleright$ if $m>j$, necessary $p<l$. Then the largest
exponent of $z_3$
 in $E(P)$ is $\leq 2j+1<2m$.\\
$\triangleright$ if $m<j$, necessary $p>l$. Then the only terms of
$E(P)$ which have $z_1t_1$ with the exponent $1$ are
\begin{center}
$\begin{array}{rcl}
L_1 & = & 8jz_1t_1y_1^{2j+2l}z_3^{2j}z_2^{2l} \\
L_2 & = & 8lz_1t_1y_1^{2j+2l}z_3^{2l}z_2^{2j} \\
L_3 & = & -4z_1t_1y_1^{2j+2}z_3^{2j}z_2^{2}\ \ \textmd{if}\ l=1 \\
L_4 & = & -4z_1t_1y_1^{2j+2}z_3^{2}z_2^{2j}\ \ \textmd{if}\ l=1. \\
\end{array}$
\end{center}
$L_1$ and $L_3$ are not colinear to $M_{m,p}$, because they are
colinear to $M_{j,l}$.\\
$L_2$ is colinear to $M_{m,p}$ only if $m=l$ and $j=p$. But then
$m=l\leq j=p\leq m$, hence $m=j$ and $p=l$, which is absurd. The
same argument may be applied to $L_4$.\\

\noindent $\bullet$ If $m+p+q=j+l$, with $1\leq q\leq p\leq m$,
then
$M_{m,p,q}=z_1t_1y_1^{2m+2q}z_3^{2m+2p}z_2^{2}t_1^{2p}z_1^{2q-2}$
does not appear in $E(P)$. In fact,\\
$\triangleright$ If $m+p>j$, the greatest exponent of $z_3$ in
$E(P)$ is $\leq 2j+1<2m+2p$.\\
$\triangleright$ If $m+p<j$, we have necessarily $q\geq l+1$, thus
$2q-1>2l\geq 2$. Then the only terms of $E(P)$ which have $z_1$
with the exponent $2q-1$ are
\begin{center}
$\begin{array}{rcl}
L_1 & = & -2C_{2j}^{2q-1}y_1^{2q+2l}z_1^{2q-1}z_2^{2j-2q+2}z_3^{2l}t_1^{2j-2q+1} \\
L_2 & = & -2C_{2j}^{2q-1}y_1^{2q}z_1^{2q-1}z_2^{2j-2q+2}z_3^{2l}t_1^{2j-2q+2l+1} \\
L_3 & = & -2C_{2j}^{2q-1}y_1^{2q}z_1^{2q-1}z_2^{2l}z_3^{2j-2q+2}t_1^{2j-2q+2l-1} \\
L_4 & = & -2C_{2j}^{2q-1}y_1^{2q+2l}z_1^{2q-1}z_2^{2l}z_3^{2j-2q+2}t_1^{2j-2q+1}. \\
\end{array}$
\end{center}
If $L_1$ is colinear to $M_{m,p,q}$, we have
$\left\{\begin{array}{rcl}
  m+q & = & q+l \\
  2 & = & 2j-2q+2 \\
  2m+2p & = & 2l \\
  2p+1 & = & 2j-2q+1 \\
\end{array}\right.$, hence $p=0$, which is absurd.\\
Similarly, $L_2,\ L_3$ and $L_4$ are not colinear to
$M_{m,p,q}$.\\
$\triangleright$ If $m+p=j$, we have necessarily $q=l$ (but only $2q-1\geq 1$). Then we have two cases:\\
$\diamond$ If $l\geq 2$, i.e. $q\geq 2$, in addition to the terms
$L_1,\dots,\ L_4$, we obtain the terms
\begin{center}
$\begin{array}{rcl}
L_5 & = & -4lz_2y_1z_3^{2j}t_1^{2j}z_1^{2l-1}y_1^{2l-1}t_1z_2 \\
L_6 & = & -4lz_2y_1z_1^{2l-1}y_1^{2l-1}t_1z_2y_1^{2j}z_3^{2j} \\
L_7 & = & -4lz_3y_1z_2^{2j}y_1^{2j}z_3t_1y_1^{2l-1}z_1^{2l-1} \\
L_8 & = & -4lz_3y_1z_3t_1y_1^{2l-1}z_1^{2l-1}t_1^{2j}z_2^{2j}.
\end{array}$
\end{center}
If $L_5$ is colinear to $M_{m,p,q}$, we have in particular $2m+2l=2l$, hence $m=0$, which is absurd.\\
Similarly, $L_6,\ L_7$ and $L_8$ are not colinear to
$M_{m,p,q}$.\\
$\diamond$ If $l=1$, i.e. $q=1$, in addition to the terms
$L_1,\dots,\ L_8$, we also obtain the terms
\begin{center}
$\begin{array}{rcl}
z_1y_1z_3^{2j}z_2^2(y_1+t_1)^{2j+2}-z_1y_1z_3^{2j}z_2^2(-y_1+t_1)^{2j+2}\\
z_1y_1z_3^{2}z_2^{2j}(y_1+t_1)^{2j+2}-z_1y_1z_3^{2}z_2^{2j}(-y_1+t_1)^{2j+2}
\end{array}$
\end{center}
Among these terms, the ones which have $t_1$ with the exponent
$2p$ vanish
because of the signs in the expansion of the binomial.\\

\textbf{Case 3.}  $P=R_3(X^{2j}Y^{2l}Z^{2k})$
(with $1\leq l\leq k\leq j$):\\
$\bullet$ According to formula (\ref{equagraphe3}), the greatest
exponent of $z_3$ in $E'_3(P)$ is $\leq 2l+2k$. So if $m=j+k+l$,
then $M_m=z_1t_1y_1^{2m}z_3^{2m}$ does not appear in
$E'_3(P)$.\\

\noindent $\bullet$ If $m+p=j+l+k$ with $1\leq p\leq m$, $m\geq
2$, then $M_{m,p}=z_1t_1y_1^{2m+2p}z_3^{2m}z_2^{2p}$ doesn't appear in $E'_3(P)$. In fact,\\
$\triangleright$ If $j+k<m$, then $z_3$ never appears in
$E'_3(P)$ with the exponent $2m$.\\
$\triangleright$ If $j+k\geq m$, we write the terms of $E'_3(P)$
which have $y_1$ with the exponent $2m+2p$:
\begin{center}
$\begin{array}{rcl|rcl}
L_1 & = & -4jz_2y_1z_1^{2j-1}y_1^{2j-1}t_1z_2y_1^{2l}z_3^{2l+2k}t_1^{2k} &  L_7 & = & -4lz_2y_1z_1^{2l-1}y_1^{2l-1}t_1z_2y_1^{2j}z_3^{2j+2k}t_1^{2k}\\
L_2 & = & -4kz_3y_1t_1^{2j}z_2^{2j+2l}y_1^{2l}z_3t_1y_1^{2k-1}z_1^{2k-1} &  L_8 & = & -4kz_3y_1t_1^{2l}z_2^{2l+2j}y_1^{2j}z_3t_1y_1^{2k-1}z_1^{2k-1}\\
L_3 & = & -4jz_2y_1z_1^{2j-1}y_1^{2j-1}t_1z_2y_1^{2k}z_3^{2k+2l}t_1^{2l} &  L_9 & = & -4kz_2y_1z_1^{2k-1}y_1^{2k-1}t_1z_2y_1^{2l}z_3^{2l+2j}t_1^{2j}\\
L_4 & = & -4lz_3y_1t_1^{2j}z_2^{2j+2k}y_1^{2k}z_3t_1y_1^{2l-1}z_1^{2l-1} &  L_{10} & = & -4jz_3y_1t_1^{2k}z_2^{2k+2l}y_1^{2l}z_3t_1y_1^{2j-1}z_1^{2j-1}\\
L_5 & = & -4kz_2y_1z_1^{2k-1}y_1^{2k-1}t_1z_2y_1^{2j}z_3^{2j+2l}t_1^{2l} &  L_{11} & = & -4lz_2y_1z_1^{2l-1}y_1^{2l-1}t_1z_2y_1^{2k}z_3^{2k+2j}t_1^{2j}\\
L_6 & = & -4lz_3y_1t_1^{2k}z_2^{2k+2j}y_1^{2j}z_3t_1y_1^{2l-1}z_1^{2l-1} &  L_{12} & = & -4jz_3y_1t_1^{2l}z_2^{2l+2k}y_1^{2k}z_3t_1y_1^{2j-1}z_1^{2j-1}.\\
\end{array}$
\end{center}
None of these terms is colinear to $M_{m,p}$, because of the exponent of $t_1$ which is always too large.\\

\noindent $\bullet$ If $m+p+q=j+l+k$ with $1\leq q\leq p\leq m$
and $(m,\,p,\,q)\neq (j,\,k,\,l)$, then
$M_{m,p,q}=z_1t_1y_1^{2m+2q}z_3^{2m+2p}z_2^{2}t_1^{2p}z_1^{2q-2}$
does not appear in $E'_3(P)$. In fact,\\
$\triangleright$ If $m+p>j+k$, then $z_3$ never appears in
$E'_3(P)$ with the exponent $2m+2p$.\\
$\triangleright$ If $m+p\leq j+k$, we write the terms of $E'_3(P)$
which have $y_1$ with exponent $2m+2q$ and $z_2$ with exponent
$2$, by setting $C_n^\delta=0$ if $\delta\notin \llbr 0,\,2k
\rrbr$:
\begin{center}
$\begin{array}{rcl|rcl}
L_1 & = & -4jz_2y_1z_1^{2j-1}y_1^{2j-1}t_1z_2y_1^{2l}z_3^{2l+2k}t_1^{2k} &  L_7 & = & -4lz_2y_1z_1^{2l-1}y_1^{2l-1}t_1z_2y_1^{2j}z_3^{2j+2k}t_1^{2k}\\
L_2 & = & -2C_{2k}^\alpha z_3y_1t_1^{2j}z_2^{2j+2l}y_1^{2l}z_3^{2k-\alpha}t_1^{2k-\alpha}y_1^{\alpha}z_1^{\alpha} &  L_8 & = & -2C_{2k}^\alpha z_3y_1t_1^{2l}z_2^{2l+2j}y_1^{2j}z_3^{2k-\alpha}t_1^{2k-\alpha}y_1^{\alpha}z_1^{\alpha}\\
L_3 & = & -4jz_2y_1z_1^{2j-1}y_1^{2j-1}t_1z_2y_1^{2k}z_3^{2k+2l}t_1^{2l} &  L_9 & = & -4kz_2y_1z_1^{2k-1}y_1^{2k-1}t_1z_2y_1^{2l}z_3^{2l+2j}t_1^{2j}\\
L_4 & = & -2C_{2l}^\beta z_3y_1t_1^{2j}z_2^{2j+2k}y_1^{2k}z_3^{2l-\beta}t_1^{2l-\beta}y_1^{\beta}z_1^{\beta} &  L_{10} & = & -2C_{2j}^\gamma z_3y_1t_1^{2k}z_2^{2k+2l}y_1^{2l}z_3^{2j-\gamma}t_1^{2j-\gamma}y_1^{\gamma}z_1^{\gamma}\\
L_5 & = & -4kz_2y_1z_1^{2k-1}y_1^{2k-1}t_1z_2y_1^{2j}z_3^{2j+2l}t_1^{2l} &  L_{11} & = & -4lz_2y_1z_1^{2l-1}y_1^{2l-1}t_1z_2y_1^{2k}z_3^{2k+2j}t_1^{2j}\\
L_6 & = & -2C_{2l}^\beta z_3y_1t_1^{2k}z_2^{2k+2j}y_1^{2j}z_3^{2l-\beta}t_1^{2l-\beta}y_1^{\beta}z_1^{\beta} &  L_{12} & = & -2C_{2j}^\gamma z_3y_1t_1^{2l}z_2^{2l+2k}y_1^{2k}z_3^{2j-\gamma}t_1^{2j-\gamma}y_1^{\gamma}z_1^{\gamma},\\
\end{array}$
\end{center}
with $\alpha\in\llbr 0,\,2k \rrbr,\ \beta\in\llbr 0,\,2l \rrbr,\
\gamma\in\llbr 0,\,2j \rrbr$ and
$\alpha,\ \beta,\ \gamma$ odd.\\
None of these terms is colinear to $M_{m,p}$, because for each of
the twelve terms, the equality of the multi-degrees gives a linear
system which leads to an absurdity. For example, if $L_1$ is
colinear to $M_{m,p}$, we have $\left\{\begin{array}{rcl}
  2m+2q & = & 2j+2l \\
  2m+2p & = & 2l+2k \\
  2p+1 & = & 1+2k \\
  2q-1 & = & 2j-1 \\
\end{array}\right.$, hence $p=k,\ q=j,\ m=l$, which is absurd
by hypothesis.\\

\textbf{Case 4.}  $P=R_3(X^{2i+1}Y^{2j+1}Z^{2k+1})$ (with $0\leq i< j< k$):\\
In this case, because of the degree, the only monomial for which
we must show that it does not appear in $E'_3(P)$ is
$\widetilde{M}_{m,p,q}$ with $m+p+q=j+k+i$, $0<q<p<m$ and
$(m,\,p,\,q)\neq (k,\,j,\,i)$. The study is the same as the one
made in the third point of case 3.\\

\noindent So we have proved the following  result$\ $:\\

\begin{Pro}\label{homolB3}$\\$
The dimension of the $0-$th space of Poisson homology of $B_3$ is
$3$, i.e.
\fbox{$\dim(HP_0(B_3))=3$}.\\
This dimension coincides with $\dim(HH_0(B_3))$.
\end{Pro}$\\$

\noindent As for the vectors of highest weight $0$ of degree
congruent to $2$ modulo $4$, we obtain recursively a formula which
enables us to express them explicitly as sums of
brackets: this is the subject of proposition~\ref{vphpdeg24}.\\

\begin{Pro}\label{vphpdeg24}$\\$
The vectors of highest weight $0$ of degree congruent to $2$
modulo $4$ are sums of brackets.
\end{Pro}

\underline{Proof:}\\
$\bullet$ We have the formula
\begin{equation}\label{et1}
\{R_3(X^{2p}),\
R_3(X^{2q})\}=\displaystyle{\frac{8}{3}pq\,R_3\big(X^{2p-1}Y^{2q-1}Z\big)}
\end{equation}
Indeed,
\begin{center}
\begin{small}
$\begin{array}{rcl} \{R_3(X^{2p}),\ R_3(X^{2q})\} & = &
\displaystyle{\frac{1}{|\go{S}_3|}\{\sum_{\sigma\in\go{S}_3}(x_{\sigma
1}y_{\sigma 2}-y_{\sigma 1}x_{\sigma 2})^{2p},\
\sum_{\tau\in\go{S}_3}(x_{\tau 1}y_{\tau 2}-y_{\tau 1}x_{\tau
2})^{2q}\}}\\
 & = & \displaystyle{\frac{1}{9}\sum_{\sigma,\ \tau\in\langle(1,2,3)\rangle}\{(x_{\sigma
1}y_{\sigma 2}-y_{\sigma 1}x_{\sigma 2})^{2p},\ (x_{\tau 1}y_{\tau
2}-y_{\tau 1}x_{\tau 2})^{2q}\}}\\
 & = & \displaystyle{\frac{4}{9}pq\sum_{\sigma,\ \tau\in\langle(1,2,3)\rangle}(x_{\sigma
1}y_{\sigma 2}-y_{\sigma 1}x_{\sigma 2})^{2p-1}(x_{\tau 1}y_{\tau
2}-y_{\tau 1}x_{\tau 2})^{2q-1}\{x_{\sigma 1}y_{\sigma
2}-y_{\sigma 1}x_{\sigma 2},\ x_{\tau 1}y_{\tau
2}-y_{\tau 1}x_{\tau 2}\}}\\
 & = & \displaystyle{\frac{4}{9}pq\sum_{\tau\in\go{S}_3}\sigma\cdot(x_1y_2-y_1x_2)^{2p-1}(x_2y_3-y_2x_3)^{2q-1}(x_3y_1-y_3x_1)}\\
 & = & \displaystyle{\frac{8}{3}pq\,R_3\big((x_1y_2-y_1x_2)^{2p-1}(x_2y_3-y_2x_3)^{2q-1}(x_3y_1-y_3x_1)\big),}
\end{array}$
\end{small}
\end{center}
where the $4-$th equality results from the table
\begin{scriptsize}
\begin{tabular}{|c||c|c|c|} \hline
  $\{\cdot\}$ & $X$ & $Y$ & $Z$ \\ \hline\hline
  $X$ & $0$ & $Z$ & $-Y$ \\ \hline
  $Y$ & $-Z$ & $0$ & $X$ \\ \hline
  $Z$ & $Y$ & $-X$ & $0$ \\ \hline
\end{tabular}
\end{scriptsize}.
\\We also have
\begin{equation}\label{et2}
R_3\big(X^{2p-1}Y^{2q-1}Z\big)R_3\big(X^{2r}\big)=\frac{1}{3}\Big[R_3\big(X^{2p+2r-1}Y^{2q-1}Z\big)+R_3\big(X^{2p-1}Y^{2q+2r-1}Z\big)
+R_3\big(X^{2p-1}Y^{2q-1}Z^{2r+1}\big)\Big]
\end{equation}
So, according to Leibniz property and formula (\ref{et1}), we have
\begin{equation}\label{et3}
\begin{array}{rcl} \{R_3(X^{2p}),\ R_3(X^{2q})R_3(X^{2r})\} & = &
\{R_3(X^{2p}),\ R_3(X^{2q})\}R_3(X^{2r})+\{R_3(X^{2p}),\
R_3(X^{2r})\}R_3(X^{2q})\\
 & = &
 \frac{8}{3}pq\,R_3\big(X^{2p-1}Y^{2q-1}Z\big)R_3\big(X^{2r}\big)+ \frac{8}{3}pr\,R_3\big(X^{2p-1}Y^{2r-1}Z\big)R_3\big(X^{2q}\big)
\end{array}
\end{equation}
$\bullet$ We deduce
\begin{equation}\label{et4}
\begin{small}
\begin{array}{rl} & R_3(X^{2p-1}Y^{2q-1}Z^{2r+1})\\
 = & 3R_3(X^{2p-1}Y^{2q-1}Z)R_3(X^{2r})-R_3(X^{2p+2r-1}Y^{2q-1}Z)-R_3(X^{2p-1}Y^{2q+2r-1}Z)\\
 = &
 \frac{9}{8pq}\{R_3(X^{2p}),\ R_3(X^{2q})R_3(X^{2r})\}-
 \frac{3r}{q}R_3(X^{2p-1}Y^{2q-1}Z)R_3(X^{2q})-R_3(X^{2p+2r-1}Y^{2q-1}Z)-R_3(X^{2p-1}Y^{2q+2r-1}Z)\\
  = & \frac{9}{8pq}\{R_3(X^{2p}),\ R_3(X^{2q})R_3(X^{2r})\}-
 \frac{r}{q}\Big[R_3(X^{2p+2q-1}Y^{2r-1}Z)+R_3(X^{2p-1}Y^{2r+2q-1}Z)+R_3(X^{2p-1}Y^{2r-1}Z^{2q+1})\Big]\\
  & -R_3(X^{2p+2r-1}Y^{2q-1}Z)-R_3(X^{2p-1}Y^{2q+2r-1}Z),
\end{array}
\end{small}
\end{equation}
where the equalities 1, 2, 3 result respectively from formulae
(\ref{et2}), (\ref{et3}), (\ref{et2}).\\
$\bullet$ Then we proceed by recurrence on $r$: we show that for
every $r\geq 1$ fixed, for every $(p,\,q)\in(\mathbb{N}^*)^2$ such
that $r<q<p$, the element $R_3(X^{2p-1}Y^{2q-1}Z^{2r-1})$ is a sum
of brackets: in fact, the case $r=1$ comes from (\ref{et1}) while
the way from $r$ to $r+1$ results from (\ref{et4}).$\blacksquare$

\subsection{\textsf{Study of $D_3$}}

\noindent The result for $D_3$ may be deduced from the result for $B_3$.\\
In fact, the vectors of highest weight $0$ are the same as the
ones of $B_3$ (this results for example from Proposition \ref{lienBnDn}), and they are given by remark~\ref{vphpB3}.\\
The element $1$ is not a bracket, according to Proposition \ref{deg0&4}.\\
Equation (\ref{equaetingofsubs}) is the same as for $B_3$. This
shows that the solutions of equation (\ref{equaetingofter})
are to be searched in the vector space $\langle R_3(X^2),\ R_3(X^2Y^2)\rangle$.\\
But these two polynomials have distinct degrees and we verify with
the help of \textsf{Maple} that none of them is a solution of
equation (\ref{equaetingofter}): see section 4.\\

\noindent So we have the following result$\ $:\\

\begin{Pro}\label{homolD3}$\\$
The dimension of the $0-$th space of Poisson homology of $D_3$ is
$1$, i.e. \fbox{$\dim(HP_0(D_3))=1$}.\\
This dimension coincides with $\dim(HH_0(D_3))$.\\
\end{Pro}

\begin{Rq}$\\$
Propositions \ref{proB2}, \ref{proD2}, \ref{homolB3} and
\ref{homolD3} show that the conjecture of J. Alev holds in the cases $B_2$, $D_2=A_1\times A_1$, $B_3$ and $D_3=A_3$.\\
\end{Rq}

\begin{Rq}$\\$
J. Alev conjectured that the equality
$\dim\,HP_0(W)=\dim\,HH_0(W)$ holds not only for the Weyl groups
of semi-simple finite-dimensional Lie algebras, but also for the
more general case of wreath products
of she form $W=\Gamma\smile \go{S}_n$, where $\Gamma$ is a finite subgroup of $\mathbf{SL}_2\mathbb{C}$. In particular:\\
$\bullet$ If $\Gamma=A_1$, we have $W=\go{S}_n$. In this case, an
explicit calculation shows that $\dim\,HP_0(W)=0=\dim\,HH_0(W)$.\\
$\bullet$ If $\Gamma=A_2$, we have $W\simeq B_n$. The cases $n=2$
and $n=3$ have been verified (\cite{AF06} and the present
article). For the case $n\geq 4$,
this is still a conjecture !\\
\end{Rq}

\section{\textsf{Formal computations}}

\noindent We collect in this section some
verifications which are carried out with \textsf{Maple}.\\

\subsection{\textsf{Definitions}}

\noindent $\bullet$ The function \verb"Image" calculates the image
of the polynomial $P\in \mathbb{A}[\mathbf{X}]$ by the matrix
$J\in \mathbf{Gl}_n\mathbb{C}$, according to the diagonal action,
with $\mathbf{X}=(\mathbf{x},\ \mathbf{y})$, and
$\mathbb{A}=\mathbb{C}$ or $\mathbb{A}=\mathbb{C}[\mathbf{z},\
\mathbf{t}]$.
\begin{verbatim}
Image:=proc(P,X::list,J) local n: n:=nops(X)/2:
subs({seq(X[i]=add(J[i,j]*X[j],j=1..n),i=1..n),seq(X[i]=add(J[i-n,j-n]*X[j],
j=n+1..2*n),i=n+1..2*n)},P); end proc:
\end{verbatim}\noindent $\bullet$ The function \verb"Reynolds" calculates the
image by the Reynolds operator associated to the group $W\subset
\mathbf{Gl}_n\mathbb{C}$ of the polynomial $P\in
\mathbb{A}[\mathbf{X}]$, according to the diagonal action.
\begin{verbatim}
Reynolds:=proc(P,X::list,W::list) local card,n: card:=nops(W):
n:=nops(X)/2: 1/card*add(Image(P,X,W[j]),j=1..card); end proc:
\end{verbatim}
\noindent $\bullet$ The function \verb"kron" gives the Kronecker
symbol of $(i,\,j)$.
\begin{verbatim}
kron:=proc(i,j) if i=j then 1 else 0 fi; end proc:
\end{verbatim}
\noindent $\bullet$ The function \verb"repr" gets a permutation
$\sigma\in\go{S}_n$, written in the shape of a list
$[\sigma(1),\dots,\ \sigma(n)]$, and gives the matrix of size $n$
and of general term $\delta_{i,\sigma(j)}$, i.e. the permutation
matrix associated to $\sigma$.
\begin{verbatim}
repr:=proc(sigma) local n: n:=nops(sigma):
Matrix(n,(i,j)->kron(i,sigma[j])); end proc:
\end{verbatim}
\noindent $\bullet$ Definition of the Weyl groups of $B_2$ and
$D_2$:
\begin{verbatim}
t12:=<<0,1>|<1,0>>: s1:=<<-1,0>|<0,1>>: s2:=<<1,0>|<0,-1>>:
s12:=<<-1,0>|<0,-1>>:
B2:=[seq(seq(seq(t12^i.s1^j.s2^k,i=0..1),j=0..1),k=0..1)]:
D2:=[seq(seq(t12^i.s12^j,i=0..1),j=0..1)]:
\end{verbatim}
\noindent $\bullet$ Definition of the Weyl group of $B_3$:
\begin{verbatim}
with(group): S3gr:=permgroup(3,{[[1,2]],[[1,2,3]]});
S3grbis:=[op(elements(S3gr))]:
S3grliste:=map(x->convert(x,'permlist',3),S3grbis):
S3:=map(repr,S3grliste): s1:=<<-1,0,0>|<0,1,0>|<0,0,1>>:
s2:=<<1,0,0>|<0,-1,0>|<0,0,1>>: s3:=<<1,0,0>|<0,1,0>|<0,0,-1>>:
B3:=[seq(seq(seq(seq(S3[i].s1^j.s2^k.s3^l,i=1..6),j=0..1),k=0..1),l=0..1)]:
\end{verbatim}
\noindent $\bullet$ Definition of the Weyl group of $D_3$:
\begin{verbatim}
s12:=<<-1,0,0>|<0,-1,0>|<0,0,1>>:s23:=<<1,0,0>|<0,-1,0>|<0,0,-1>>:
D3:=[seq(seq(seq(S3[i].s12^j.s23^k,i=1..6),j=0..1),k=0..1)]:
\end{verbatim}

\subsection{\textsf{Verification of the calculations of propositions \ref{proB2} and \ref{proD2}}}

\noindent The following calculations enable us to verify that
$E_2(R_2(X^2))=0$ for $B_2$ and $E_2(R_2(X^2))\neq 0$ for $D_2$.\\
\begin{verbatim}
ind:=[seq(x[j],j=1..2),seq(y[j],j=1..2)]:
\end{verbatim}
\begin{verbatim}
uB:=Reynolds((x[1]*y[2]-y[1]*x[2])^2,ind,B2);
vB:=add(z[j]*y[j]-t[j]*x[j],j=1..2)*subs({seq(x[j]=x[j]+z[j],j=1..2),
seq(y[j]=y[j]+t[j],j=1..2)},uB):
equaB:=Reynolds(vB,ind,B2):equaBd:=expand(equaB);
\end{verbatim}
\begin{verbatim}
uD:=Reynolds((x[1]*y[2]-y[1]*x[2])^2,ind,D2);
vD:=add(z[j]*y[j]-t[j]*x[j],j=1..2)*subs({seq(x[j]=x[j]+z[j],j=1..2),
seq(y[j]=y[j]+t[j],j=1..2)},uD):
equaD:=Reynolds(vD,ind,D2):equaDd:=expand(equaD);
\end{verbatim}

\subsection{\textsf{Verification of the identities of Proposition \ref{deg0&4&8}}}

\noindent The following calculations prove the identities
$E_3(R_3(X^2))=0$ and $E_3(R_3(X^2Y^2))=0$.\\
\begin{verbatim}
ind:=[seq(x[j],j=1..3),seq(y[j],j=1..3)]:
\end{verbatim}
\begin{verbatim}
u1:=Reynolds((x[1]*y[2]-y[1]*x[2])^2,ind,B3):
v1:=add(z[j]*y[j]-t[j]*x[j],j=1..3)*subs({seq(x[j]=x[j]+z[j],j=1..3),
seq(y[j]=y[j]+t[j],j=1..3)},u1):
equa1:=Reynolds(v1,ind,B3):equad1:=expand(equa1);
\end{verbatim}

\begin{verbatim}
u2:=Reynolds((x[1]*y[2]-y[1]*x[2])^2*(x[2]*y[3]-y[2]*x[3])^2,ind,B3):
v2:=add(z[j]*y[j]-t[j]*x[j],j=1..3)*subs({seq(x[j]=x[j]+z[j],j=1..3),
seq(y[j]=y[j]+t[j],j=1..3)},u2):
equa2:=Reynolds(v2,ind,B3):equad2:=expand(equa2);
\end{verbatim}

\subsection{\textsf{Verification of the calculations of Proposition \ref{homolD3}}}

\noindent The following calculations show that the polynomials
$R_3(X^2)$ and $R_3(X^2Y^2)$ are not solutions of equation (\ref{equaetingofter}).\\
\begin{verbatim}
ind:=[seq(x[j],j=1..3),seq(y[j],j=1..3)]:
\end{verbatim}
\begin{verbatim}
u1:=Reynolds((x[1]*y[2]-y[1]*x[2])^2,ind,D3):
v1:=add(z[j]*y[j]-t[j]*x[j],j=1..3)*subs({seq(x[j]=x[j]+z[j],j=1..3),
seq(y[j]=y[j]+t[j],j=1..3)},u1):
equa1:=Reynolds(v1,ind,D3):equad1:=expand(equa1): nops(equad1);
\end{verbatim}

\begin{verbatim}
u2:=Reynolds((x[1]*y[2]-y[1]*x[2])^2*(x[2]*y[3]-y[2]*x[3])^2,ind,D3):
v2:=add(z[j]*y[j]-t[j]*x[j],j=1..3)*subs({seq(x[j]=x[j]+z[j],j=1..3),
seq(y[j]=y[j]+t[j],j=1..3)},u2):
equa2:=Reynolds(v2,ind,D3):equad2:=expand(equa2): nops(equad2);
\end{verbatim}

\begin{Large}
\noindent\\ \textbf{Acknowledgements} \end{Large}\\

\noindent I would like to thank my thesis advisors Gadi Perets and
Claude Roger for their efficient and likeable help, for their
great availability, and for the
time that they devoted to me all along this study.\\
I also thank Jacques Alev for a fruitful
discussion.\\
And I thank Serge Parmentier who read over my English text.\\


\begin{thebibliography}{}
\bibitem[AF06]{AF06}
Jacques Alev and Lo\"{\i}c Foissy, \textit{Le groupe des traces de
Poisson de la vari\'et\'e quotient $\go{h}\oplus\go{h}^*/W$ en
rang
$2$}.\\
\bibitem[AFLS00]{AFLS00}
Jacques Alev, Marco A. Farinati, Thierry Lambre and Andrea L.
Solotar, \textit{Homologie des invariants d'une alg\`ebre de Weyl
sous l'action d'un
 groupe fini}, Journal of Algebra, Volume 232, pp. 564-577, 2000.\\
\bibitem[AL98]{AL98}
Jacques Alev and Thierry Lambre, \textit{Comparaison de
l'homologie de Hochschild et de l'homologie de Poisson pour une
d\'eformation des surfaces de Klein}. In Algebra and operator
theory
(Tashkent, 1997), pp. 25-38. Kluwer Acad. Publ., Dordrecht, 1998.\\
\bibitem[AL99]{AL99}
Jacques Alev and Thierry Lambre, \textit{Homologie des invariants
d'une alg\`ebre de Weyl}, K-Theory, Volume 18, pp. 401-411, 1999.\\
\bibitem[B88]{B88}
Jean-Luc Brylinsky, \textit{A differential complex for Poisson manifolds},
J. Differential Geometry 28, pp. 93-114, 1988.\\
\bibitem[BEG04]{BEG04} Yuri Berest, Pavel Etingof and Victor
Ginzburg, \textit{Morita equivalence of Cherednik algebras}, J.
Reine Angew. Math.
\textbf{568}, 81-98, 2004.\\
\bibitem[C72]{C72} R. W. Carter, \textit{Conjugacy classes in the Weyl group},
Compositio Mathematica,
Vol. \textbf{25}, Fasc. 1, 1-59, 1972.\\
\bibitem[DCP76]{DCP76}
C. De Concini and C. Procesi, \textit{A characteristic-free
approach to invariant theory}, Adv. Math.
\textbf{21}, 330-354, 1976.\\
\bibitem[EG02]{EG02}
Pavel Etingof and Victor Ginzburg, \textit{Symplectic reflection
algebras, Calogero-Moser Space, and deformed Harish-Chandra
homomorphism}, Invent. Math.
\textbf{147}, no. 2, pp. 243-348, 2002.\\
\bibitem[FH91]{FH91}
William Fulton and Joe Harris, \textit{Representation Theory: A First Course }, Springer-Verlag, 1991.\\
\bibitem[GK04]{GK04}
Sudhir R. Ghorpade and Christian Krattenthaler, \textit{The
Hilbert Serie of Pfaffien Rings}, Algebra, arithmetic and geometry
with
applications (West Lafayette, IN, 2000), 337--356, Springer, Berlin, 2004.\\
\bibitem[GRS07]{GRS07}
Laurent Guieu and Claude Roger, with an appendix from Vlad
Sergiescu, \textit{L'Alg\`ebre et le Groupe de Virasoro: aspects
g\'eom\'etriques et alg\'ebriques, g\'en\'eralisations},
Publication du Centre de Recherches Math\'ematiques de Montr\'eal,
s\'erie ``Monographies, notes
de cours et Actes de conf\'erences'', PM28, 2007.\\
\bibitem[H72]{H72}
James E. Humphreys, \textit{Representations of semisimple Lie
algebras}, Springer-Verlag, 1972.\\
\bibitem[R02]{R02}
Hannah Robbins, \textit{Invariant and Covariant Rings of Finite
Pseudo-Reflection Groups}, 2002.\\
\bibitem[Sp77]{Sp77}
T. A. Springer, \textit{Invariant theory},
 Lecture Notes in Math., \textbf{585}, Springer-Verlag, 1977.\\
\bibitem[St93]{St93}
Bernd Sturmfels, \textit{Algorithms in Invariant Theory},
 Texts and Monographs in Symbolic Computation, Springer-Verlag, 1993.\\
\end{thebibliography}
\end{document}